\documentclass[lettersize,journal]{IEEEtran}

\usepackage{amsmath,amssymb,amsthm}
\usepackage{graphicx}
\graphicspath{{figs/}}
\usepackage[caption=false,font=footnotesize]{subfig}
\usepackage{booktabs,array}
\usepackage{algorithm}
\usepackage{algpseudocode}
\usepackage{enumitem}
\usepackage{xcolor}
\usepackage{pifont}
\usepackage{url}
\usepackage{hyperref}
\usepackage{cleveref}

\newtheorem{theorem}{Theorem}
\newtheorem{lemma}[theorem]{Lemma}
\newtheorem{proposition}[theorem]{Proposition}
\newtheorem{definition}[theorem]{Definition}

\newtheorem{observation}[theorem]{Observation}

\IEEEoverridecommandlockouts

\begin{document}

\title{\textsf{PLATRS}: A Lattice-Based Linkable Accountable Ring Signature for Permissioned Blockchains}

\author{
	Jie Zhang, Xiaohong Li,
	Honglin Mao,
	Ruitao Feng,~\IEEEmembership{Member,~IEEE},
	Lingyun Yuan, \\
	and Guangdong Bai,~\IEEEmembership{Member,~IEEE}
}

\markboth{IEEE Transactions on Information Forensics and Security}{Zhang \MakeLowercase{\textit{et al.}}: Linkable Accountable Ring Signatures for Permissioned Blockchains}

\maketitle
\begin{abstract}
	Cross-domain data sharing on permissioned ledgers requires both anonymity and accountability: an endorsement should hide its author, while a regulator must be able to identify the author when the same authorization is used twice. Existing lattice-based solutions address these properties separately: publicly traceable schemes reveal the signer to anyone who detects a collision, linkable schemes without tracing cannot open a signature, and threshold constructions answer a different question. Accountable lattice ring signatures already exist, so the combination is not the contribution; what they leave open is a per-event tag whose noise is keyed to the tracing secret, and an opening that one authority performs and any verifier rechecks.
	
	We propose \textsf{PLATRS}, a module-lattice linkable accountable ring signature for permissioned blockchains. Members are already enrolled by the consortium certificate authority, so the key an opening returns is bound to a registered identity. The signer chooses the ring, verification is index-free, and each signature carries a per-event tag that any verifier can compare. The tag noise is derived from the signer's tracing secret and the event, so that linking is a single comparison for honestly generated signatures while the tag remains an M-LWE sample without the secret. Detection is two-tier: a tag mismatch is detected immediately, and when two accepted signatures on one event carry different tags, the ledger triggers an opening; the authority publishes the public opening component it recovers, and any member recomputes the verdict against the ring. We prove CPA anonymity from M-LWE and linkability, authority-only traceability, and unforgeability from M-SIS with M-LWE in the random oracle model. At $N=1024$, our open-source implementation signs in 9.886~ms, verifies in 3.919~ms, links in 0.7~$\mu$s, and opens in 480~$\mu$s, with a 107.3~KiB signature that grows logarithmically with the ring size; adjudication, which re-verifies at the recovered index, takes 4.867~ms. A Hyperledger Fabric deployment sustains 60.5\ signature submissions per second at 7.87~MB/s with $16$~MiB blocks, a rate set by commit bandwidth rather than by arithmetic or cores. Under a common harness it is the fastest of the 5\ measured schemes to sign and to verify and second on signature size, and against the operation-count model of FLLL$^+$21 it signs 3188$\times$ faster with a 133$\times$ shorter signature.
\end{abstract}

\textbf{
	\textcolor{blue!70!black}{
		This work has been submitted to the IEEE for possible publication. Copyright may be transferred without notice, after which this version may no longer be accessible.
	}
}

\section{Introduction}
\label{sec:introduction}

Cross-organizational data exchange is routine in clinical research, inter-bank settlement, and multi-tier manufacturing. A record produced by one institution is consumed and audited by another, and permissioned blockchains have become the trust anchor for such flows~\cite{DBLP:journals/csur/ZhangXL19,DBLP:conf/tase/MaoZZL25}, including bilateral trades that must settle payment and data together~\cite{PriMEDeal}. A permissioned chain also settles the question that makes accountability possible at all: every participant is enrolled, since the membership service provider binds each public key to a legal identity and consensus admits only enrolled organisations. An opening that returns a public key therefore names a member, without an external registry. This setting creates a direct conflict between anonymity and accountability. Providers need anonymity because the existence of a query may reveal which hospital holds a rare-disease cohort, which bank is screening a client, or which supplier is short of capacity. Regulators, on the other hand, need accountability because an anonymous channel that cannot be audited may be abused for money laundering or collusion~\cite{ref26,MediChainAudit}. The question is therefore one of digital forensics as much as of cryptography: how a consortium detects misuse publicly, names the responsible member only when a dispute exists, and lets a third party recheck that naming.

Ring signatures~\cite{ref8} provide anonymity, and their linkable and traceable variants~\cite{ref14,ref114} provide different forms of accountability by embedding a tag that can be compared across signatures. Most deployed constructions rely on discrete logarithms or pairings~\cite{ref45} and are therefore vulnerable to Shor's algorithm~\cite{DBLP:journals/siamcomp/Shor97}. Their lattice-based replacements~\cite{ref49} are post-quantum, but currently offer a weaker accountability model than the classical schemes they aim to replace.

Post-quantum ring signatures are already practical. At $N=1024$, the baselines driven in Section~\ref{sec:performance} sign a ring in 43.34~ms to 700581.1~ms, and \textsf{PLATRS} signs in 9.886~ms. The main limitation appears when accountability is required. FLLL$^+$21~\cite{FLLL+21} and YLGT23~\cite{YLGT23} trace publicly, so a collision reveals the signer to every node that observes it; CXGG$^+$24~\cite{FS-LLRS} links but cannot open; JESS$^+$26~\cite{jagganath2026lotrs} addresses threshold endorsement with neither linking nor tracing; DZWW$^+$24~\cite{DZWW+24} is compact but pairing-based; accountable ring signatures~\cite{Bootle15} open publicly but do not link; and lattice group signatures with message-dependent opening~\cite{LMN16} need a second authority and a per-message token. In summary, the schemes above do not key their tag noise to the tracing secret, and none keeps the opening with a single authority while letting any verifier recheck it. The combination is no longer novel in itself: SPRING~\cite{ZhangLuoHuang2026} builds an accountable lattice ring signature by sign-then-prove and observes that a tag-to-key relation can be added to it. What remains open is the tag's derivation, because a publicly computable noise admits tag recovery.

The deployment setting imposes three additional constraints. First, the tag must be derived per event; otherwise, a shared tag would link a member's signatures across unrelated collaborations. Second, verification must be index-free; otherwise, any node could recover the signer's index by running $N$ trial verifications. Third, opening must be confined to one authority, and the authority must publish a proof that the ciphertext opens to the key it recovers, so that the verdict can be publicly recomputed.

To address these requirements, we propose \textsf{PLATRS}, a linkable accountable ring signature over module lattices. Setup publishes two module matrices and the authority's encryption key. Each member registers a signing key and, separately, a public opening component, so that authenticating and being named remain different powers. To endorse an event, the member derives a matrix from the event identifier, forms the tag $\boldsymbol{\tau}=\mathbf{A}_{\mathsf{event}}\mathbf{s}_2+\mathbf{e}_{\mathsf{tag}}$ with the noise expanded from the opening secret and the event, encrypts the public opening component to the authority, and proves in one proof that it owns a ring member's keys, that the tag carries short noise, and that the ciphertext contains the matching key. Verification requires no index, linking is a tag comparison, and opening is one decryption followed by a key match. The tag relation fixes the noise as short, and linkability is defined over signatures produced by $\mathsf{Sign}$. When two accepted signatures on one event carry different tags, the contract releases the opening, the authority publishes the public opening component it decrypts together with a proof that the ciphertext carries it, and anyone rechecks the verdict against the ring. Attribution thus rests on M-SIS and M-LWE rather than on how a client was implemented.

\noindent\textbf{Contributions.}
The main contributions are summarized as follows.
\begin{itemize}[leftmargin=10px]
	\item \emph{Model and construction.} We formalize a lattice-based linkable accountable ring signature model that simultaneously captures a signer-chosen ring, public per-event linking, and authority-only opening, and we give a module-lattice construction, \textsf{PLATRS}, that realizes it. In this construction, the condition for opening is enforced by ledger state rather than by a second authority, so public linkability and authority-only traceability are reconciled within one primitive (Section~\ref{sec:system}).
	
	\item \emph{Keyed tag and two-tier detection.} A central design principle of \textsf{PLATRS} is to decouple public linkability from public traceability by keying the tag noise to the signer's tracing secret and the event. This yields deterministic per-event tags for honestly generated signatures, closes the tag-recovery channel left open by public rounding, and keeps the tag pseudorandom under M-LWE without the tracing secret (Section~\ref{sec:tag-attack}). Building on this tag, we introduce a two-tier detection rule that combines an $O(1)$ tag comparison with a ledger-triggered opening; the authority publishes the recovered key with a proof of correct decryption, so the verdict is publicly recomputable and the authority cannot frame a member (Section~\ref{sec:link-alg}).
	
	\item \emph{Security and evaluation.} We establish CPA anonymity from M-LWE and linkability, authority-only traceability, and unforgeability from M-SIS with M-LWE in the random oracle model, with the games and reductions given in Appendix~B (Section~\ref{sec:security}). We further provide an open-source implementation and a Hyperledger Fabric integration, and conduct a capability-conditioned comparison that rebuilds five published schemes under one harness, covering signing, verification, linking, opening, and signature size (Section~\ref{sec:performance}).
\end{itemize}
%The remainder of this paper is organized as follows. Section~\ref{sec:related} reviews related work. Section~\ref{sec:system} presents the system overview and security definitions. Section~\ref{sec:prelim} introduces preliminaries. Section~\ref{sec:construction} describes the construction. Section~\ref{sec:security} analyzes security. Section~\ref{sec:performance} reports the implementation and evaluation. Section~\ref{sec:limitations} discusses limitations, and Section~\ref{sec:conclusion} concludes the paper.

\begin{figure*}[htbp]
	\centering
	\includegraphics[width=\textwidth]{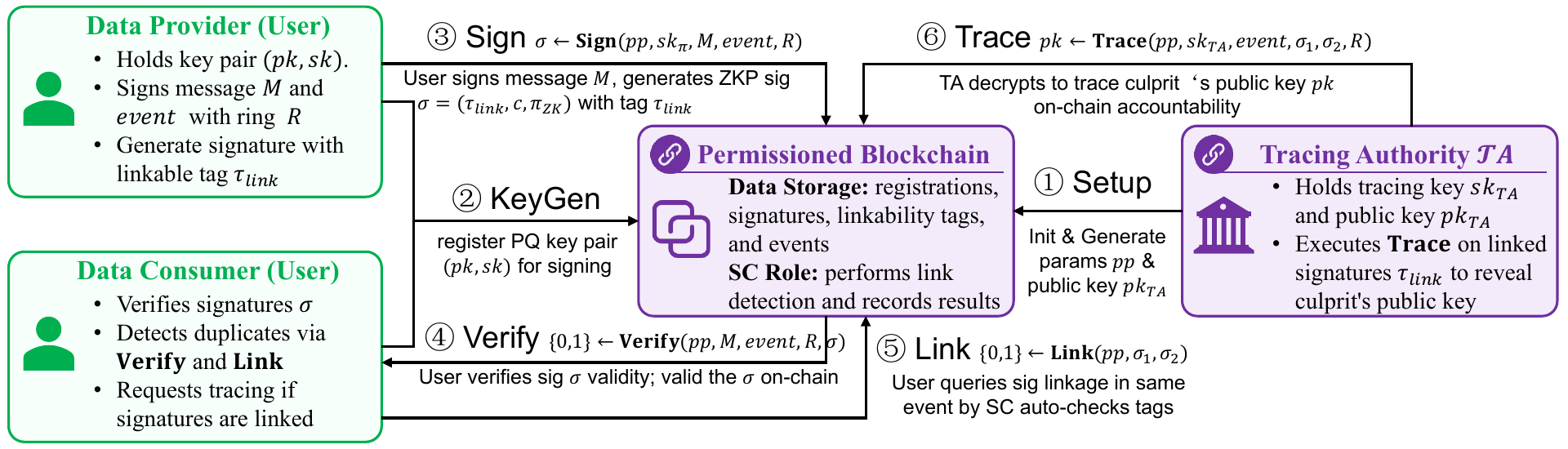}
	\caption{PLATRS workflow: \textbf{1.} Setup, $\mathcal{TA}\to$ ledger: $(\mathsf{pp},\mathbf{P}_{\mathsf{TA}})\leftarrow\mathsf{Setup}(1^\lambda)$; \textbf{2.} KeyGen, user $\to$ ledger: $\mathsf{pk}\leftarrow\mathsf{KeyGen}(\mathsf{pp})$; \textbf{3.} Sign, provider $\to$ ledger: $\sigma=(\boldsymbol{\tau}_{\mathsf{link}},\mathbf{c},\pi_{\mathsf{alg}},\pi_{\mathsf{sel}})\leftarrow\mathsf{Sign}(\mathsf{sk}_\pi,M,\mathsf{event},R)$; \textbf{4.} Verify, ledger $\to$ consumer; \textbf{5.} Link, consumer: $\mathsf{Link}(\sigma_1,\sigma_2)$; \textbf{6.} Open/Judge, $\mathcal{TA}$: $\mathsf{Open}(\mathsf{sk}_{\mathsf{TA}},\mathsf{event},\sigma)\to\mathsf{pk}^{\ast}$.}
	\label{fig2}
\end{figure*}

\section{Related Work}
\label{sec:related}

\subsection{Accountable Data Sharing and Ring Signatures}
\label{sec:related-ads}

Accountable data-sharing systems decide what may be shared and who may audit it, from auditable cohort models~\cite{ZHANGTDSC35} and verifiable fuzzy search~\cite{zhang2025dvfs} to cross-domain sharing over a federated ledger~\cite{DBLP:conf/tase/MaoZZL25} and quantified anonymity with attribute-based access~\cite{zhang2025qaebac}, with trust and reputation deciding which partners may share at all~\cite{IJISFan}. They authenticate by identifier or credential, so privacy comes from encrypting payloads rather than from hiding the actor; \textsf{PLATRS} supplies the missing token, one anonymous signature per access event.

Ring signatures~\cite{ref8} hide a signer in a self-chosen set and have identity-based~\cite{ref17} and revocable~\cite{ref114} variants. Linkable schemes let anyone test whether two signatures share a signer~\cite{ref10}, enough for double-spend detection but revealing that a participant's signatures are related; traceable schemes open a signature always~\cite{ref14,ref15} or after a threshold~\cite{DZWW+24}. What matters in deployment is who can open, because public traceability turns every collision into a deanonymization~\cite{ref26}.

\subsection{Accountable Opening and $k$-Times Authentication}
\label{sec:related-ars}

An opening authority and an anonymous signer are compatible once the primitive carries an opening key: accountable ring signatures let the signer name an opener beside the ring, and a public $\mathsf{Judge}$ checks the opening, so an opener cannot accuse a member without evidence. Bootle~et al. reaches $O(\log N)$ size through the one-out-of-many argument~\cite{Bootle15} that our ring also uses, and the lattice construction of Libert~et al.~\cite{LLMNW16} is the closest post-quantum relative of our opening step.

A second axis is how often a member may act: $k$-times anonymous authentication bounds the count per period at a cost independent of $k$~\cite{TS06}, with lattice instantiations under a Stern-style argument~\cite{Song25}, ledger nullifiers~\cite{BCGMMZ14}, or a threshold opening in Lk-TRS~\cite{Xie25}.
Opening can also be conditional: Libert, Mouhartem and Nguyen tie it to a per-message token released by a second authority~\cite{LMN16}, a Stern-era reference point rather than a runtime baseline. \textsf{PLATRS} keeps the condition in the ledger instead, the contract deciding when an opening may be requested and the recovered key letting anyone recompute the verdict.

\subsection{Post-Quantum Ring Signatures and Baselines}
\label{sec:related-pq}

Lattice assumptions~\cite{ref58} are a main post-quantum replacement for discrete logarithms and pairings~\cite{ref45}, which quantum algorithms break~\cite{DBLP:journals/siamcomp/Shor97}; implementations are surveyed in~\cite{ref49}. Practical lattice ring signatures have been built from one-out-of-many proofs~\cite{DBLP:conf/acns/EsginSSLZ19}, DualRing-style compression and SMILE-style set membership~\cite{DBLP:conf/crypto/LyubashevskyNS21}, accumulators~\cite{DBLP:journals/joc/LibertLNW23}, and unique or threshold variants~\cite{ref32,jagganath2026lotrs}.

We compare with five published schemes, rebuilt under a common harness rather than quoted: FLLL$^+$21~\cite{FLLL+21}, a framework for traceable ring signatures with public tracing; YLGT23~\cite{YLGT23}, a lattice traceable ring signature on a TripleRing structure, also public; CXGG$^+$24~\cite{FS-LLRS}, forward-secure linking without tracing; DZWW$^+$24~\cite{DZWW+24}, a RingCT protocol from a linkable threshold ring signature over pairings, hence not post-quantum; and JESS$^+$26~\cite{jagganath2026lotrs}, a structured threshold ring signature with neither linking nor tracing. Classical traceable and revocable schemes~\cite{ref14,ref15,ref114,Wen2025Revocable} are discrete-logarithm based and not competitors under a quantum adversary.

Two axes separate \textsf{PLATRS} from every entry we evaluate that supports opening: the ring is chosen by the signer, and opening is confined to one authority while linking remains public; FLLL$^+$21 and YLGT23 open to anyone who detects a collision, CXGG$^+$24 and JESS$^+$26 do not open, and no entry separates the two acts.
\subsection{Concurrent Lattice Accountability}
\label{sec:related-concurrent}

Recent lattice constructions approach the same conflict from different sides. ZLH26~\cite{ZhangLuoHuang2026} builds a ring signature by sign-then-prove, a BLISS-style signature with a zero-knowledge set-membership argument, and makes it compact and extensible; it observes that a tag-to-key relation can be added, so the tag is an extension rather than a built-in event identifier. HLYH$^+$26~\cite{ZebraCPA} targets vehicular authentication with a lattice traceable ring signature over a consortium chain, deliberately removing linkability tags and embedding a tracing trapdoor that only authorised auditors recover, with a threshold-tracing extension; detection there is a tracing operation, not a public tag comparison. DY26~\cite{DBLP:journals/cybersec/DongY26} stays with group signatures: a manager registers the members, implicit, explicit and sequential linkability are placed under user control, key-oblivious encryption limits the opener, and a polynomial mechanism handles revocation. None of the three keys its tag noise to the tracing secret, and none combines a public per-event tag with an opening that a single authority performs and any verifier rechecks.

\begin{table*}[t]
\centering
\caption{The closest lattice constructions. ``Extension'' means the tag relation is sketched rather than built in; ``conditional'' means opening follows a policy rather than a dispute record.}
\label{tab:concurrent}
\footnotesize
\setlength{\tabcolsep}{3pt}
\begin{tabular}{@{}p{2.3cm}p{5.2cm}p{3.3cm}p{4.1cm}@{}}
\toprule
Scheme & Construction & Public link & Opening \\
\midrule
\textsf{PLATRS} & module-lattice ring, signer-chosen ring & per-event tag & one TA, publicly rechecked \\
ZLH26~\cite{ZhangLuoHuang2026} & lattice ring, sign-then-prove & extension & designated opener \\
HLYH$^+$26 \cite{ZebraCPA} & lattice TRS with ZK & none (tags removed) & auditor trapdoor (threshold) \\
DY26~\cite{DBLP:journals/cybersec/DongY26} & lattice group signature & user-controlled, sequential & conditional, key-oblivious \\
\bottomrule
\end{tabular}
\end{table*}

\section{System Overview}
\label{sec:system}

\textsf{PLATRS} runs over a permissioned ledger and provides post-quantum accountable anonymity for cross-domain data sharing~\cite{DBLP:journals/tifs/LvFMZWL24}.

\subsection{Entities}
\label{sec:entities}

Three roles interact as depicted in Fig.~\ref{fig2}.
A \emph{user} holds a lattice key pair and signs for a ring without revealing which member it is~\cite{DBLP:conf/cscwd/ZhangLZHBF25}.
A \emph{tracing authority} $\mathcal{TA}$, an honest-but-curious auditor or threshold committee~\cite{DBLP:journals/tdsc/LinHNH23}, holds a decryption key and opens a signature only when the ledger records a conflict.
The \emph{ledger} stores registrations, signatures, tags, openings and verdicts, and its contracts compare tags to flag double-endorsement without revealing identity~\cite{ZHANGTDSC35,zhang2025qaebac}.

\noindent\textbf{Identity comes from the chain.}
A ring is a set of \emph{enrolled} keys: the consortium CA issues a credential to every organisation the channel admits, and registration records the mapping from that credential to a lattice key.
The primitive therefore opens to a public key and stops there; resolving that key to a legal entity is the registry lookup the chain already supports, which is why authority-only opening suffices for a consortium and would not suffice on a permissionless chain.
Table~\ref{tab:notation} collects the symbols used below.

\begin{table}[t]
\centering
\caption{Notation.}
\label{tab:notation}
\footnotesize
\setlength{\tabcolsep}{3pt}
\begin{tabular}{@{}l@{\hspace{6pt}}l@{}}
\toprule
$\mathcal{R}_q$ & $\mathbb{Z}_q[X]/(X^d+1)$, $d$ a power of two \\
$\mathbf{A},\mathbf{B}$ & module matrices in $\mathcal{R}_q^{k\times\ell}$ \\
$\mathbf{s}_1,\mathbf{s}_2$ & signing and opening secrets \\
$\mathbf{P}_1,\mathbf{P}_2$ & $\mathbf{A}\mathbf{s}_1+\mathbf{e}_1$ and $\mathbf{B}\mathbf{s}_2+\mathbf{e}_2$ \\
$R$, $N$, $\pi$ & ring, its size, signer index \\
$\kappa$ & $\lceil\log_{b_{\mathsf{sel}}} N\rceil$ selector digits \\
$\mathbf{M},\mathbf{u},\mathbf{w}$ & proof matrix, statement, witness \\
$\beta$, $\eta$ & M-SIS norm bound, noise bound \\
$\mathsf{event}$ & authorisation identifier \\
$\mathbf{A}_{\mathsf{event}}$ & matrix derived from $\mathsf{event}$ \\
$\boldsymbol{\tau}_{\mathsf{link}}$ & per-event tag \\
$\mathbf{e}_{\mathsf{tag}}$ & tag noise, $\mathsf{Expand}(H_2(\mathbf{s}_2\|\mathsf{event}))$ \\
$\mathbf{c}$ & ciphertext of $\mathbf{P}_2$ to $\mathcal{TA}$ \\
$\pi_{\mathsf{alg}}$, $\pi_{\mathsf{sel}}$ & algebraic and selector proofs \\
\bottomrule
\end{tabular}
\end{table}

\subsection{Ledger Rules}
\label{sec:ledger}

The algorithms decide predicates; the ledger decides when each is asked and records the answer.
A record is keyed by $(\mathsf{event},R)$ and moves through three states.

\noindent\textbf{Endorse.}
A submitted $\sigma$ is admitted iff $\mathsf{Verify}(\mathsf{pp},M,\mathsf{event},R,\sigma)=1$.
On the first admission the contract stores $\boldsymbol{\tau}_{\mathsf{link}}$ and the record becomes \textsc{endorsed}.
A second submission carrying the same tag is rejected as a duplicate of the first: this is the $O(1)$ check every node runs, and it stops duplicate use by any client that executes $\mathsf{Sign}$.
A second submission carrying a different tag is admitted, the record becomes \textsc{conflict}, and the contract emits a dispute.

\noindent\textbf{Dispute and opening.}
The contract admits an opening request only for a record in \textsc{conflict}, a rule enforced by the contract and recorded on the ledger.
$\mathcal{TA}$ calls $\mathsf{Open}$ separately on both records of the conflict and submits the two public opening components with their proofs of correct decryption; the contract runs $\mathsf{Judge}$ on each, places each key in $R$, and stores both verdicts, so the conflict is resolved by comparing two openings rather than by the tag.
Two consequences follow from putting the opening on-chain rather than in a channel: its target and its count are public, so an authority that opened signatures nobody disputed leaves a permanent trace; and the verdict is recomputed by the contract from the published key, so a false claim is rejected rather than believed.
What the ledger does \emph{not} do is keep the authority out: $\mathcal{TA}$ holds $\mathsf{sk}_{\mathsf{TA}}$ and can decrypt any ciphertext it is given, so confining opening to disputes is a rule the contract enforces and records, not a property the ciphertext enforces.

\noindent\textbf{Security implications of the two-tier design.}
Tier~1 is preventive and free, but it assumes the client executes $\mathsf{Sign}$; tier~2 is detective and costs one opening per conflict, yet it assumes nothing about the client, because $\mathsf{Judge}$ compares the opened public component rather than the derivation that produced the tag.
If the two openings name different members, the conflict is two members each using one authorisation, not one member using two.

\subsection{Protocol}
\label{sec:proto-overview}

Eight polynomial-time algorithms (Section~\ref{sec:construction}).
$\mathsf{Setup}(1^\lambda)$ outputs $\mathsf{pp}$ with $\mathbf{P}_{\mathsf{TA}}$ and $\mathbf{A},\mathbf{B}$, plus $\mathsf{sk}_{\mathsf{TA}}=\mathbf{S}_{\mathsf{TA}}$; $\mathsf{KeyGen}$ gives $\mathsf{pk}=(\mathbf{P}_1,\mathbf{P}_2)$ and $\mathsf{sk}=(\mathbf{s}_1,\mathbf{s}_2,\mathbf{e}_1,\mathbf{e}_2)$.
$\mathsf{Sign}(\mathsf{sk}_\pi,M,\mathsf{event},R)$ returns $(\sigma,\boldsymbol{\varrho})$ with $\sigma=(\boldsymbol{\tau}_{\mathsf{link}},\mathbf{c}_1,\mathbf{c}_2,\hat{\mathbf{u}},\pi_{\mathsf{alg}},\pi_{\mathsf{sel}})$ and opening $\boldsymbol{\varrho}$, whose two-layer proof attests that the signer belongs to $R$, that $\boldsymbol{\tau}_{\mathsf{link}}$ is correctly formed with short noise $\|\mathbf{e}_{\mathsf{tag}}\|_\infty\le\eta$, and that $(\mathbf{c}_1,\mathbf{c}_2)$ encrypts the matching $\mathbf{P}_{2,\pi}$; the signature carries $\hat{\mathbf{u}}$ explicitly, so that a verifier never needs an index to rebuild the instance.
$\mathsf{Verify}$ decides from $\sigma$ alone whether it is a valid signature of $R$ on $(M,\mathsf{event})$, with no index and no secret.
$\boldsymbol{\varrho}$ is the opening of $\hat{\mathbf{u}}$ and is never part of $\sigma$; $\mathsf{Sign}$ hands it to the signer alone.
$\mathsf{Confirm}$ takes an index $\pi$, the signature and that opening, and returns one bit: because $\hat{\mathbf{u}}$ is hiding, an index test needs $\boldsymbol{\varrho}$, so the $N$ trial runs that would recover $\pi$ are not available to anyone else.
$\mathsf{Link}$ compares tags on the same event; $\mathsf{Open}$ decrypts $\mathbf{c}$ under $\mathsf{sk}_{\mathsf{TA}}$ returns the public opening component carried inside it, and proves that the ciphertext opens to that component; $\mathsf{Judge}$ verifies the proof and places the key in $R$, so an opening is a claim that anyone rechecks rather than a statement that must be believed.

\subsection{Deployment Scenarios}
\label{sec:scenarios}

The three constraints of Section~\ref{sec:introduction} fix what the tag may span; what remains is who uses it.
In clinical cohort sharing the ledger rejects a second endorsement of an access token while the regulator opens only a clinician flagged for repeated access; in inter-bank audit duplicate settlement is caught by any node and the counterparty disclosed only to the supervisor; in supply-chain attestation a duplicate attestation for one purchase order is detectable while cross-order anonymity hides the sub-tier structure.

\subsection{Threat Model}
\label{sec:threat}

All games of Section~\ref{sec:security} share one experiment. $\mathsf{Setup}$ runs, $\mathsf{pp}$ is public, $\mathsf{sk}_{\mathsf{TA}}$ is held by the authority, and $H_1$--$H_3$ together with $\mathsf{Expand}$ are random oracles, with $H_1$ deriving $\mathbf{A}_{\mathsf{event}}$, $H_2$ the tag seed, $H_3$ the Fiat--Shamir challenge and $\mathsf{Expand}$ the tag noise.
The adversary $\mathcal{A}$ is PPT, learns the public keys of the members of a ring $R$, and fixes a set $C\subset R$ of statically corrupted members with $|C|\le N-2$; corrupted members deviate arbitrarily, honest members follow the algorithms.
$\mathcal{A}$ has adaptive access to $\mathcal{O}_{\mathsf{Sign}}(i,M,\mathsf{event},R)$ for $i\notin C$, to $\mathsf{Verify}$ on any signature, and to each random oracle; $\mathsf{Confirm}$ is available with any index and any candidate opening, the strongest form of the query, which answers nothing about $\sigma$ without the opening $\boldsymbol{\varrho}$ that only the signer holds. on-chain data is public, consensus is secure~\cite{TRMRaft}, and the ledger orders but never rewrites a finalized entry.
$\mathcal{TA}$ is honest-but-curious: it follows $\mathsf{Open}$, never leaks $\mathsf{sk}_{\mathsf{TA}}$ and does not collude with $\mathcal{A}$; an opening oracle is available only for a signature the ledger has flagged, which models the publication of an attestation rather than an unrestricted opening oracle.
Signing is Fiat--Shamir with aborts, and the restart count is visible in timing.
Beyond the random oracle model we assume $\mathsf{M\text{-}LWE}$, $\mathsf{M\text{-}SIS}$ and IND-CPA security of $\Pi_{\mathsf{TA}}$; the sigma protocol of Section~\ref{sec:zkpok} is simulated, not assumed: its zero knowledge is statistical and its simulation extractability is Proposition~\ref{prop:sim-ext}.

\subsection{Security Definitions}
\label{sec:sec-def}

We fix the syntax and correctness of the primitive first, then the games.
Every game is played in the random oracle model: $\mathcal{C}$ runs $\mathsf{Setup}$, hands $\mathsf{pp}$ to a PPT adversary $\mathcal{A}$, answers signing queries through $\mathcal{O}_{\mathsf{Sign}}$, and lets $\mathcal{A}$ corrupt users statically; an advantage is negligible in $\lambda$ unless stated otherwise.
Two modelling choices delimit the claims: linkability is a property of signatures produced by $\mathsf{Sign}$ (Definition~\ref{def:link-honest}), and the behaviour of a client that departs from it is handled by the ledger rather than by a game (Section~\ref{sec:ledger}); traceability is \emph{authority-only}, constraining who may open a signature rather than when, since $\mathsf{sk}_{\mathsf{TA}}$ decrypts any ciphertext and the dispute rule is an interface of an honest authority.
Rings consist of honestly generated keys.

\begin{definition}[Syntax]
\label{def:syntax}
A \emph{publicly verifiable, publicly linkable, authority-only traceable ring signature} over messages $\mathcal{M}$ and events $\mathcal{E}$ is a tuple $\Pi=(\mathsf{Setup},\mathsf{KeyGen},\mathsf{Sign},\mathsf{Verify},\mathsf{Confirm},\mathsf{Link},\mathsf{Open},\mathsf{Judge})$ of polynomial-time algorithms with the signatures
\begin{align*}
\mathsf{Setup}(1^\lambda)&\to(\mathsf{pp},\mathsf{sk}_{\mathsf{TA}}),\\
\mathsf{KeyGen}(\mathsf{pp})&\to(\mathsf{pk},\mathsf{sk}),\\
\mathsf{Sign}(\mathsf{pp},\allowbreak\mathsf{sk},\allowbreak M,\allowbreak\mathsf{event},\allowbreak R)&\to(\sigma,\boldsymbol{\varrho})\text{ or }\bot,\\
\mathsf{Verify}(\mathsf{pp},\allowbreak M,\allowbreak\mathsf{event},\allowbreak R,\allowbreak\sigma)&\to\{0,1\},\\
\mathsf{Confirm}(\mathsf{pp},\allowbreak M,\allowbreak\mathsf{event},\allowbreak R,\allowbreak i,\allowbreak\sigma,\allowbreak\boldsymbol{\varrho})&\to\{0,1\},\\
\mathsf{Link}(\mathsf{pp},\allowbreak\mathsf{event},\allowbreak\sigma_1,\allowbreak\sigma_2)&\to\{\mathsf{linked},\mathsf{independent}\},\\
\mathsf{Open}(\mathsf{pp},\allowbreak\mathsf{sk}_{\mathsf{TA}},\allowbreak\mathsf{event},\allowbreak\sigma)&\to(\mathsf{pk}^{\ast},\pi_{\mathsf{dec}})\text{ or }\bot,\\
\mathsf{Judge}(\mathsf{pp},\allowbreak M,\allowbreak\mathsf{event},\allowbreak R,\allowbreak\sigma,\allowbreak\mathsf{pk}^{\ast},\allowbreak\pi_{\mathsf{dec}})&\to\{\mathsf{convict},\mathsf{acquit}\},
\end{align*}
where $M\in\mathcal{M}$, $\mathsf{event}\in\mathcal{E}$, $R=(\mathsf{pk}_1,\dots,\mathsf{pk}_N)$ is a ring, $i\in[N]$, and $\pi_{\mathsf{dec}}$ is the authority's proof that $\mathbf{c}$ opens to $\mathsf{pk}^{\ast}$ under $\mathbf{P}_{\mathsf{TA}}$ (Section~\ref{sec:link-alg}); the ciphertext randomness is a witness of that proof and is never returned.
\end{definition}

\noindent\textbf{Verification, attribution, and the opening.}
$\mathsf{Verify}$ is the ordinary public predicate: no index, no secret, the same input for every verifier, deciding whether $\sigma$ is well formed on $(M,\mathsf{event},R)$: the selection layer accepts and the algebraic layer proves relation~\eqref{eq:relation} over the committed pair, the tag and the ciphertext.  Whether the committed pair is a ring entry is the narrower claim that $\mathsf{Confirm}$ decides at a named index and $\mathsf{Judge}$ re-checks at the opened one.
$\mathsf{Confirm}$ takes the index together with the opening $\boldsymbol{\varrho}$ and decides the narrower claim ``member $i$ signed'', which is what the signer runs on its own signature and what turns an attestation into evidence, the opening being the signer's own data rather than something a verifier can supply.
$\mathsf{Open}$ is the only algorithm that consumes $\mathsf{sk}_{\mathsf{TA}}$, and $\mathsf{Judge}$ is public: it checks the decryption proof, places the key in $R$, and decides whether $\sigma$ is by the member that key names.
A party holding neither $i$ nor $\mathsf{sk}_{\mathsf{TA}}$ learns from $\sigma$ only what $\mathsf{Link}$ reveals, namely \emph{that} a well-formed record exists and never \emph{who}; Proposition~\ref{prop:index-hiding} states that neither predicate reveals $\pi$.

\begin{definition}[Correctness]
\label{def:correct}
$\Pi$ is \emph{correct} if for every $\lambda$, every $(\mathsf{pp},\mathsf{sk}_{\mathsf{TA}})\leftarrow\mathsf{Setup}(1^\lambda)$, every $(\mathsf{pk},\mathsf{sk})\leftarrow\mathsf{KeyGen}(\mathsf{pp})$, every $M\in\mathcal{M}$, every $\mathsf{event}\in\mathcal{E}$ and every ring $R$ that contains $\mathsf{pk}$ at position $i$, the following hold with probability $1-\mathsf{negl}(\lambda)$ over the coins of $\Pi$, for $(\sigma,\boldsymbol{\varrho})\leftarrow\mathsf{Sign}(\mathsf{pp},\mathsf{sk},M,\mathsf{event},R)$ and any $\sigma_1,\sigma_2$ so obtained:
\begin{enumerate}[leftmargin=16px,itemsep=0.1em,topsep=0.2em,parsep=0pt]
\item \emph{verification}: $\mathsf{Verify}(\mathsf{pp},M,\mathsf{event},R,\sigma)=1$ and $\mathsf{Confirm}(\mathsf{pp},M,\mathsf{event},R,i,\sigma,\boldsymbol{\varrho})=1$;
\item \emph{linking}: $\mathsf{Link}(\mathsf{pp},\mathsf{event},\sigma_1,\sigma_2)=\mathsf{linked}$;
\item \emph{opening}: $(\mathsf{pk},\pi_{\mathsf{dec}})\leftarrow\mathsf{Open}(\mathsf{pp},\mathsf{sk}_{\mathsf{TA}},\mathsf{event},\sigma)$ and $\mathsf{Judge}(\mathsf{pp},M,\mathsf{event},R,\sigma,\mathsf{pk},\pi_{\mathsf{dec}})=\mathsf{convict}$, for $\sigma\leftarrow\mathsf{Sign}(\mathsf{pp},\mathsf{sk},M,\mathsf{event},R)$.
\end{enumerate}
\end{definition}

\begin{definition}[CPA anonymity]
\label{def:anon}
$\mathcal{A}$ outputs $M^\ast$, $\mathsf{event}^\ast$, a ring $R^\ast$ of honest users and $i_0,i_1\in R^\ast$ that it has neither corrupted nor queried on $(\cdot,\mathsf{event}^\ast,R^\ast)$; it may call $\mathsf{Verify}(\cdot,\cdot,\cdot,R^\ast,\cdot)$ on the challenge as often as it likes, since the public predicate answers only whether \emph{some} member signed, it may call $\mathsf{Confirm}$ with any index and any candidate opening, and it is not given the opening $\boldsymbol{\varrho}$.
It receives $\sigma^\ast\leftarrow\mathsf{Sign}(\mathsf{pp},\mathsf{sk}_{i_b},M^\ast,\mathsf{event}^\ast,R^\ast)$ for $b\leftarrow\{0,1\}$ and returns $b'$, winning if $b'=b$; write $\mathsf{Adv}^{\mathsf{anon}}_{\mathcal{A}}(\lambda)=|\Pr[b'=b]-1/2|$.
\end{definition}

CCA anonymity drops the restriction on queries; Section~\ref{sec:cpa-vs-cca} shows that public per-event linkability rules it out, which is why CPA is the right notion here.

\begin{definition}[Linkability for honestly generated signatures, $\mathsf{Link}_{\mathsf{hon}}$]
\label{def:link-honest}
$\mathcal{A}$ has $\mathcal{O}_{\mathsf{Sign}}$ and an opening oracle $\mathcal{O}_{\mathsf{Open}}(\mathsf{event},\sigma)$, which returns $\mathsf{Open}(\mathsf{pp},\mathsf{sk}_{\mathsf{TA}},\mathsf{event},\sigma)$ for an accepting $\sigma$ and $\bot$ otherwise.
It wins by outputting $(\mathsf{event},R,\sigma_1,\sigma_2)$ such that either
\begin{enumerate}[leftmargin=16px,itemsep=0.15em,topsep=0.25em,parsep=0pt]
\item $\sigma_1,\sigma_2$ were both returned by $\mathcal{O}_{\mathsf{Sign}}$ for one honest ring member on $\mathsf{event}$, both verify, and $\mathsf{Link}(\sigma_1,\sigma_2)=\mathsf{independent}$; or
\item both verify and $\mathsf{Link}(\sigma_1,\sigma_2)=\mathsf{linked}$, while $\mathcal{O}_{\mathsf{Open}}(\mathsf{event},\sigma_1)$ and $\mathcal{O}_{\mathsf{Open}}(\mathsf{event},\sigma_2)$ return two distinct keys, or the key of an honest user who signed $\mathsf{event}$ at most once.
\end{enumerate}
\end{definition}

Condition~(1) is the linking guarantee for the algorithms we specify, and condition~(2) is soundness of $\mathsf{Link}$: a linked pair may not be attributed to two parties, nor to an honest party that behaved.
Condition~(2) quantifies over \emph{arbitrary} accepting signatures, since it follows from the tag equation rather than from the derivation of the noise.
Condition~(1) holds for signatures that follow $\mathsf{Sign}$; a member that re-randomises its own tag is handled by the ledger rule of Section~\ref{sec:ledger}, which resolves the case at detection rather than at prevention.

\begin{definition}[Authority-only opening]
\label{def:trace}
$\mathcal{C}$ keeps $\mathsf{sk}_{\mathsf{TA}}$.
$\mathcal{A}$ wins if it outputs $(M,\mathsf{event},R,\sigma,\mathsf{pk}^{\ast},\pi_{\mathsf{dec}})$ such that $\mathsf{Verify}(\mathsf{pp},M,\mathsf{event},R,\sigma)=1$ and $\mathsf{Judge}(\mathsf{pp},M,\mathsf{event},R,\sigma,\mathsf{pk}^{\ast},\pi_{\mathsf{dec}})=\mathsf{convict}$, while no member holding $\mathsf{pk}^{\ast}$ produced $\sigma$; equivalently, no accepting signature can be opened onto a member that did not sign it.
The definition constrains what an opening may assert and says nothing about \emph{when} a ciphertext may be opened, which is why the primitive is authority-only rather than conditionally openable.
\end{definition}

\begin{definition}[Unforgeability]
\label{def:unforgeability}
$\mathcal{A}$ outputs $(M,\mathsf{event},R,i,\sigma)$ with $R$ containing only honest, uncorrupted users and $i\in R$, and wins if $\mathsf{Verify}(\mathsf{pp},M,\mathsf{event},R,\sigma)=1$ and $\mathsf{Confirm}(\mathsf{pp},M,\allowbreak\mathsf{event},\allowbreak R,\allowbreak i,\allowbreak\sigma,\allowbreak\boldsymbol{\varrho})=1$ while $(M,\mathsf{event},R,\cdot)$ was never returned by $\mathcal{O}_{\mathsf{Sign}}$ for $i$; a forger must therefore name the member it impersonates, and the index it names is the one $\mathsf{Confirm}$ checks.
\end{definition}

\section{Preliminaries}
\label{sec:prelim}

\noindent
$\lambda$ is the security parameter and all algorithms are PPT unless stated otherwise.
Work takes place in $\mathcal{R}_q=\mathbb{Z}_q[X]/(X^d+1)$ with $q$ prime and $d$ a power of two; $\|a\|_\infty$ is the largest coefficient magnitude, and $S_\eta=\{a\in\mathcal{R}_q:\|a\|_\infty\le\eta\}$.
Vectors and matrices over $\mathcal{R}_q$ are bold, e.g.\ $\mathbf{v}\in\mathcal{R}_q^k$ and $\mathbf{A}\in\mathcal{R}_q^{k\times\ell}$.

\subsection{Lattice Assumptions}
\label{sec:assump}

Security rests on two module-lattice problems~\cite{YLGT23,ref32} that underpin recent post-quantum ring and group signatures~\cite{FLLL+21}.

\begin{definition}[M-SIS$_{k,\ell,q,\beta}$]
Given $\mathbf{A}\xleftarrow{\$}\mathcal{R}_q^{k\times\ell}$, find a non-zero $\mathbf{x}\in\mathcal{R}_q^\ell$ with $\mathbf{A}\mathbf{x}=\mathbf{0}\bmod q$ and $\|\mathbf{x}\|_\infty\le\beta$. This generalises SIS to module lattices and is believed hard for quantum adversaries at suitable parameters~\cite{YLGT23,ref32}.
\end{definition}

\noindent\textbf{Augmented instances.}
Every M-SIS instance that arises below is $[\mathbf{A}\mid\mathbf{I}_k]$ with $\mathbf{A}\xleftarrow{\$}\mathcal{R}_q^{k\times\ell}$, and a solution is a pair $(\mathbf{x},\mathbf{y})\neq\mathbf{0}$ with $\mathbf{A}\mathbf{x}+\mathbf{y}=\mathbf{0}$ and $\|\mathbf{x}\|_\infty,\|\mathbf{y}\|_\infty\le\beta$; taking $\mathbf{y}=\mathbf{0}$ recovers M-SIS on $\mathbf{A}$ alone.
The identity block is not eliminated by a reduction from the bare matrix, and none is claimed: hardness is assumed for the augmented form, and the estimates of Appendix~\ref{app:params} are taken at its column count $m=d(k+\ell)$, so the block belongs to the estimated instance rather than being adjoined to it afterwards.
The same applies to $\mathbf{A}_{\mathsf{event}}$, which is programmed through $H_1$ and is uniform in the random-oracle model.

\noindent\textbf{Distributions.}
Two distributions share the support $\{-\eta,\dots,\eta\}$ and differ only in variance, which is why both are written out here once.
Sampling $\mathbf{x}\xleftarrow{\$} S_\eta^\ell$ means uniform coefficients on $\{-\eta,\dots,\eta\}$, of variance $v_S=\eta(\eta+1)/3$; it supplies the secrets $\mathbf{s}_1,\mathbf{s}_2$ and the mask $\mathbf{r}$.
The error distribution $\chi$ is the sum of $\eta$ independent uniform signs over the same support, of variance $v_\chi=\eta$, and it supplies $\mathbf{e}_1,\mathbf{e}_2,\mathbf{e}_3,\mathbf{e}_4$ and, through $\mathsf{Expand}$, the tag noise.
Every noise term in the scheme is therefore a sample of $\chi$: there is one noise law and one estimate, and no bound is transferred between distributions.

\begin{definition}[M-LWE$_{k,\ell,q,\eta}$]
Let $\chi$ be the signed-sum distribution above: polynomial coefficients supported on $\{-\eta,\dots,\eta\}$ with variance $\eta$. Given $\mathbf{A}\xleftarrow{\$}\mathcal{R}_q^{k\times\ell}$, $\mathbf{s}\xleftarrow{\$} S_\eta^\ell$ and $\mathbf{e}\xleftarrow{\$}\chi^k$, distinguish $(\mathbf{A},\mathbf{A}\mathbf{s}+\mathbf{e})$ from $(\mathbf{A},\mathbf{u})$ with $\mathbf{u}\xleftarrow{\$}\mathcal{R}_q^k$. Hardness is inherited from LWE and underpins anonymity and confidentiality here~\cite{YLGT23,ref32}.
\end{definition}

Here this error distribution gives at least $\lambda$ bits against BKZ and quantum sieving~\cite{YLGT23,ref32}.

\subsection{The Keyed Tag}
\label{sec:rounding}

Prior lattice TRS derive the tag noise from a floor quotient or from a rounding remainder of a \emph{public} value; we instead key the noise with the signer's own tracing secret, which removes the algebraic handle an adversary needs in order to strip the noise off the tag.

\noindent\textbf{Tag noise.}
Let $\chi$ be the centred binomial on $\{-\eta,\dots,\eta\}$ used for all key noise, let $H_2:\{0,1\}^*\to\{0,1\}^{\lambda_{\mathsf{seed}}}$ be a hash with $\lambda_{\mathsf{seed}}=256$ fixed independently of the selector digit count $\kappa$, and let $\mathsf{Expand}:\{0,1\}^{\lambda_{\mathsf{seed}}}\to\{-\eta,\dots,\eta\}^{k\cdot d}$ be an extendable-output expansion; both are modelled as random oracles.
On input $(\mathbf{s}_2,\mathsf{event})$ the signer derives $\mathbf{e}_{\mathsf{tag}}=\mathsf{Expand}\bigl(H_2(\mathbf{s}_2\|\mathsf{event})\bigr)\in\{-\eta,\dots,\eta\}^{k}$ by hashing a fixed encoding of $\mathbf{s}_2$, so that $\|\mathbf{e}_{\mathsf{tag}}\|_\infty\le\eta$, and sets $\boldsymbol{\tau}_{\mathsf{link}}=\mathbf{A}_{\mathsf{event}}\mathbf{s}_2+\mathbf{e}_{\mathsf{tag}}$.
Thus $\mathbf{e}_{\mathsf{tag}}$ is deterministic in $(\mathbf{s}_2,\mathsf{event})$, so one signer's repeated use of one authorisation yields one tag, and for an unused event it is uniform and independent of $\mathbf{s}_2$, so the tag is an M-LWE sample with error drawn from $\chi$-range noise (Lemma~\ref{lem:tag-prf}).
The tags revealed per tracing secret are polynomially many, so anonymity treats them as polynomially many M-LWE samples of one secret, which are jointly pseudorandom because stacking them only increases the number of samples.

\subsection{Why the Tag Noise Must Be Keyed}
\label{sec:tag-attack}

The tag is the only value compared across signatures, hence the only place where an algebraic attack replaces a lattice one.
Suppose the noise were a deterministic public function, as in a rounding construction: $\mathbf{e}_{\mathsf{tag}}=\mathsf{rem}_p(\mathbf{C}\mathbf{s}_2+\mathbf{D}_{\mathsf{event}})$ with $\mathbf{C}\in\{0,1\}^{k\times\ell}$, $\mathbf{D}_{\mathsf{event}}=H_2(\mathsf{event})$ and a public modulus $p$.
Writing $\mathbf{C}\mathbf{s}_2+\mathbf{D}_{\mathsf{event}}=p\,\mathbf{t}+\mathbf{e}_{\mathsf{tag}}$ gives
$\boldsymbol{\tau}_{\mathsf{link}}-\mathbf{D}_{\mathsf{event}}+p\,\mathbf{t}=\bigl(\mathbf{A}_{\mathsf{event}}+\mathbf{C}\bigr)\mathbf{s}_2 \pmod q$,
where $\mathbf{t}$ follows from $\mathbf{D}_{\mathsf{event}}$ alone on all but a vanishing fraction of coordinates, because $\|\mathbf{C}\mathbf{s}_2\|_\infty\ll p/2$: away from a rounding boundary the quotient is determined without any secret.
The system is then \emph{noiseless} in $\mathbf{s}_2$, and two signatures of the signer on other events, which CPA anonymity grants, supply $2kd>\ell d$ equations at Set-III, enough to solve by Gaussian elimination without touching a lattice problem.
The argument covers every public rounding map, and Theorem~\ref{thm:tag-necessity} states it in that generality.

\begin{theorem}[Public tag noise admits tag recovery]
\label{thm:tag-necessity}
Let a ring signature scheme carry a per-event tag $\boldsymbol{\tau}=\mathbf{A}_{\mathsf{event}}\mathbf{s}+\mathbf{e}$, with $\mathbf{A}_{\mathsf{event}}\in\mathcal{R}_q^{k\times\ell}$ public, $\mathbf{s}\in\mathcal{R}_q^{\ell}$ the signer's secret and $\mathbf{e}=\mathsf{Noise}(\mathsf{pp},\mathsf{event})$ a deterministic public function of $\mathsf{pp}$ and the event.
If $2k\ge\ell$ and $\bigl[\mathbf{A}_{\mathsf{ev}_1};\mathbf{A}_{\mathsf{ev}_2}\bigr]$ has rank $\ell$ over $\mathcal{R}_q$ for two events, the scheme does not satisfy CPA anonymity: an adversary recovers $\mathbf{s}$ from two signatures of the challenge users on other events and decides the challenge bit with advantage $1/2$.
\end{theorem}

\noindent\emph{Proof.}
The CPA game lets $\mathcal{A}$ query signatures of $i_0$ and $i_1$ on any event other than $\mathsf{event}^\ast$.
Query both on two events $\mathsf{ev}_1\neq\mathsf{ev}_2$, evaluate $\mathbf{e}_j=\mathsf{Noise}(\mathsf{pp},\mathsf{ev}_j)$, and subtract: $\mathbf{A}_{\mathsf{ev}_j}\mathbf{s}=\boldsymbol{\tau}_j-\mathbf{e}_j$ for $j=1,2$.
The two signatures give $2k\ge\ell$ linear equations in the $\ell$ ring elements of $\mathbf{s}$, which the rank condition solves by Gaussian elimination over $\mathcal{R}_q$; evaluating $\boldsymbol{\tau}^\ast_b=\mathbf{A}_{\mathsf{event}^\ast}\mathbf{s}_{i_b}+\mathsf{Noise}(\mathsf{pp},\mathsf{event}^\ast)$ for $b=0,1$ then identifies the challenge signer with advantage $1/2$. \hfill$\square$

\noindent
At Set-III, $2k=10$ and $\ell=6$, so two signatures suffice: keying the noise removes exactly this, because $\mathsf{Expand}(H_2(\mathbf{s}_2\|\mathsf{event}))$ is not public and the system is no longer linear in $\mathbf{s}_2$.

\noindent\textbf{Keyed noise.}
With $\mathbf{e}_{\mathsf{tag}}=\mathsf{Expand}(H_2(\mathbf{s}_2\|\mathsf{event}))$ the hypothesis of Theorem~\ref{thm:tag-necessity} fails: the noise is not $p\,\mathbf{t}+\mathbf{e}$ for a publicly computable $\mathbf{t}$, and there is no modulus against which to cancel it, so eliminating $\mathbf{e}_{\mathsf{tag}}$ from $\boldsymbol{\tau}_{\mathsf{link}}-\mathbf{A}_{\mathsf{event}}\mathbf{s}_2=\mathbf{e}_{\mathsf{tag}}$ requires a short vector in a coset, which is M-SIS; anonymity reduces to M-LWE instead (Theorem~\ref{thm:anon}).
Removing the rounding modulus from $\mathsf{pp}$ is thus the countermeasure rather than a simplification.
The cost is definitional rather than computational: \eqref{eq:rel-tag} enforces shortness, not derivation, and amounts to one expansion per signature.

\subsection{Encryption for the Tracing Authority}
\label{sec:enc}

For traceability with public hiding, $\mathcal{TA}$ alone uses an M-LWE encryption scheme $\Pi_{\mathsf{TA}}=(\mathsf{TAKeyGen},\mathsf{Enc},\mathsf{Dec})$, following Lindner--Peikert~\cite{LP11} in the module setting.
$\mathsf{TAKeyGen}$ samples $\mathbf{S}_{\mathsf{TA}}\xleftarrow{\$} S_\eta^{\ell\times\ell}$ and $\mathbf{E}_{\mathsf{TA}}\xleftarrow{\$}\chi^{k\times\ell}$ and publishes $\mathbf{P}_{\mathsf{TA}}=\mathbf{B}\mathbf{S}_{\mathsf{TA}}+\mathbf{E}_{\mathsf{TA}}$; $\mathsf{Enc}$ is the map of~\eqref{eq:tag}, and $\mathsf{Dec}$ computes $\mathbf{c}_2-\mathbf{S}_{\mathsf{TA}}^{\top}\mathbf{c}_1=\mathsf{encode}(\mathbf{P}_2)+(\mathbf{e}_4-\mathbf{S}_{\mathsf{TA}}^{\top}\mathbf{e}_3)$ and rounds, recovering $\mathbf{P}_2$ since the error stays short.
The scheme is IND-CPA under M-LWE.

\subsection{Zero-Knowledge Proof of Knowledge}
\label{sec:zkpok}

Signing emits a NIZKPoK for linear relations over $\mathcal{R}_q$ with small-norm witnesses, via Fiat--Shamir with aborts~\cite{ref58} as in recent lattice ring and group signatures~\cite{DBLP:conf/acns/EsginSSLZ19,DBLP:conf/crypto/LyubashevskyNS21}.
The prover commits to masked witnesses, receives a small-coefficient challenge $c\in\mathcal{C}$ and answers, aborting and restarting when the answer exceeds a threshold; Fiat--Shamir with $c=H_3(\mathsf{inst},M,\mathsf{com})$ makes it non-interactive~\cite{FS86}.
In the random oracle model this is complete, multi-theorem zero knowledge (a simulator programming $H_3$ exists) and simulation extractable (Proposition~\ref{prop:sim-ext}, which also fixes the forking loss), the properties used for Theorems~\ref{thm:anon}--\ref{thm:unforgeability}; we write $\pi=(\pi_{\mathsf{alg}},\pi_{\mathsf{sel}})\gets\mathsf{Prove}(\mathsf{inst},\mathsf{wit})$ for the two-layer proof.
Zero knowledge is what stops $\boldsymbol{\tau}_{\mathsf{link}}$ and $\mathbf{c}$ from revealing which member signed.

\subsection{Event Identifiers}
\label{sec:event}

As in traceable ring signatures~\cite{ref14,DBLP:conf/tase/MaoZZL25,DBLP:journals/tdsc/LinHNH23}, an explicit identifier binds a signature to one chain operation, e.g.\ $\mathsf{event}=\mathsf{SHAKE256}(\textsf{chain\_id}\|\textsf{authorization\_id}\|\textsf{resource}\|\textsf{version})$, so linkability is per event and the identifier names the authorisation rather than the transaction that carries it: a retry at another height or index still links, while a different authorisation gives an independent tag, detecting duplicate spending or access without sacrificing cross-event anonymity~\cite{zhang2025qaebac}.

\section{Construction of PLATRS}
\label{sec:construction}

\textsf{PLATRS} instantiates the model of Section~\ref{sec:proto-overview} on the primitives of Section~\ref{sec:prelim}: full anonymity from a pseudorandom tag plus a zero-knowledge proof hiding the signer's index; per-event public linkability from $\mathsf{Link}$, which compares tags, the proof binding the tag to the same opening secret that the ciphertext encrypts, so that two signatures produced by $\mathsf{Sign}$ on one event under one signer carry the identical tag and are flagged by every verifier; and authority-only traceability from $\mathsf{Open}$, which decrypts the public opening component of a linked signer and nothing else.

\subsection{Setup and Key Generation}
\label{sec:setup-alg}
\label{sec:keygen-alg}

$\mathsf{Setup}(1^\lambda)$ fixes $d$ a power of two, a prime $q\equiv1\pmod{2d}$, ranks $k,\ell$ and bounds $\eta,\beta$ such that M-SIS$_{k,\ell,q,2\beta}$ and M-LWE$_{k,\ell,q,\eta}$ reach $\lambda$ bits (Section~\ref{sec:parameters}), draws $\mathbf{A},\mathbf{B}\xleftarrow{\$}\mathcal{R}_q^{k\times\ell}$ and $\mathbf{S}_{\mathsf{TA}}\xleftarrow{\$} S_\eta^{\ell\times\ell}$, $\mathbf{E}_{\mathsf{TA}}\xleftarrow{\$}\chi^{k\times\ell}$, sets $\mathbf{P}_{\mathsf{TA}}=\mathbf{B}\mathbf{S}_{\mathsf{TA}}+\mathbf{E}_{\mathsf{TA}}$, and outputs $\mathsf{pp}=(d,q,k,\ell,\eta,\beta,\mathbf{A},\mathbf{B},\mathbf{P}_{\mathsf{TA}},H_1,H_2,H_3,\mathsf{Expand})$ with $\mathsf{sk}_{\mathsf{TA}}=\mathbf{S}_{\mathsf{TA}}$.
$\mathsf{KeyGen}$ draws $\mathbf{s}_1,\mathbf{s}_2\xleftarrow{\$} S_\eta^\ell$ and $\mathbf{e}_1,\mathbf{e}_2\xleftarrow{\$}\chi^k$, and returns $\mathsf{pk}=(\mathbf{A}\mathbf{s}_1+\mathbf{e}_1,\,\mathbf{B}\mathbf{s}_2+\mathbf{e}_2)$ and $\mathsf{sk}=(\mathbf{s}_1,\mathbf{s}_2,\mathbf{e}_1,\mathbf{e}_2)$.
Table~\ref{tab:params} lists the evaluated sets; the requirements that pin them, the concrete security estimate and the tracing-noise calibration are in Appendix~\ref{sec:parameters}.

\subsection{Sign}
\label{sec:sign-alg}

User $\pi$ holds $\mathsf{sk}_\pi=({\mathbf{s}}_{1,\pi},{\mathbf{s}}_{2,\pi},{\mathbf{e}}_{1,\pi},{\mathbf{e}}_{2,\pi})$, $\mathsf{pk}_\pi=({\mathbf{P}}_{1,\pi},{\mathbf{P}}_{2,\pi})$ and signs $M$ for $\mathsf{event}$ in $R=\{\mathsf{pk}_1,\dots,\mathsf{pk}_N\}$.

\subsubsection{Tag, ciphertext and the relation}
\label{sec:zk-instantiation}

Let $\mathbf{A}_{\mathsf{event}}=H_1(\mathsf{event})\in\mathcal{R}_q^{k\times\ell}$ and $\mathbf{e}_{\mathsf{tag}}=\mathsf{Expand}\bigl(H_2(\mathbf{s}_{2,\pi}\|\mathsf{event})\bigr)\in\{-\eta,\dots,\eta\}^{k}$. The signer forms the tag and the tracing ciphertext,
\begin{equation}\label{eq:tag}
	\small
\begin{aligned}
 \boldsymbol{\tau}_{\mathsf{link}}&=\mathbf{A}_{\mathsf{event}}\cdot{\mathbf{s}}_{2,\pi}+\mathbf{e}_{\mathsf{tag}},\\
 \mathbf{c}=(\mathbf{c}_1,\mathbf{c}_2)&=\bigl(\mathbf{B}^{\top}\mathbf{r}+\mathbf{e}_3,\ \mathbf{P}_{\mathsf{TA}}^{\top}\mathbf{r}+\mathbf{e}_4+\mathsf{encode}({\mathbf{P}}_{2,\pi})\bigr),
\end{aligned}
\end{equation}
with $\mathbf{r}\xleftarrow{\$} S_\eta^k$, $\mathbf{e}_3,\mathbf{e}_4\xleftarrow{\$}\chi^\ell$ and $\mathsf{encode}:\mathcal{R}_q^{k}\to\mathcal{R}_q^{\ell}$ injective (e.g.\ zero padding).
Both $\mathbf{e}_{\mathsf{tag}}$ and $\boldsymbol{\tau}_{\mathsf{link}}$ are determined by $({\mathbf{s}}_{2,\pi},\mathsf{event})$, which is what makes repeated use detectable, while by Lemma~\ref{lem:tag-prf} the tag is simultaneously an M-LWE sample of ${\mathbf{s}}_{2,\pi}$ across events.
The signer then proves that both are correctly formed for one ring member without revealing which. The relation $\mathcal{R}_{\mathsf{anon}}$ takes $\mathsf{inst}=(\mathbf{A},\mathbf{B},\mathbf{A}_{\mathsf{event}},\mathbf{P}_{\mathsf{TA}},R,\boldsymbol{\tau}_{\mathsf{link}},\mathbf{c})$, witness $\mathsf{wit}=(\mathbf{s}_1,\mathbf{s}_2,\mathbf{e}_1,\mathbf{e}_2,\mathbf{e}_{\mathsf{tag}},\mathbf{r},\mathbf{e}_3,\mathbf{e}_4,\pi)$ with $\pi\in[N]$, and accepts iff
\begin{align}
 {\mathbf{P}}_{1,\pi} &= \mathbf{A}\mathbf{s}_1+\mathbf{e}_1\ \land\ \|\mathbf{s}_1\|_\infty,\|\mathbf{e}_1\|_\infty\le\beta, \label{eq:rel-sig}\\
 {\mathbf{P}}_{2,\pi} &= \mathbf{B}\mathbf{s}_2+\mathbf{e}_2\ \land\ \|\mathbf{s}_2\|_\infty,\|\mathbf{e}_2\|_\infty\le\beta, \label{eq:rel-tracekey}\\
 \boldsymbol{\tau}_{\mathsf{link}} &= \mathbf{A}_{\mathsf{event}}\mathbf{s}_2+\mathbf{e}_{\mathsf{tag}}\ \land\ \|\mathbf{e}_{\mathsf{tag}}\|_\infty\le\eta, \label{eq:rel-tag}\\
 \mathbf{c}_1 &= \mathbf{B}^{\top}\mathbf{r}+\mathbf{e}_3\ \land\ \|\mathbf{e}_3\|_\infty\le\beta, \label{eq:rel-c1}\\
 \mathbf{c}_2 &= \mathbf{P}_{\mathsf{TA}}^{\top}\mathbf{r}+\mathbf{e}_4+\mathsf{encode}({\mathbf{P}}_{2,\pi})\ \land\ \|\mathbf{e}_4\|_\infty\le\beta. \label{eq:rel-c2}
\end{align}
Equation~\eqref{eq:rel-tag} fixes the tag as a short-noise image of $(\mathbf{s}_2,\mathsf{event})$ with noise bounded by $\eta$, and this is the only place where signer honesty could be exploited: a proof of knowledge establishes that \emph{some} short $\mathbf{e}_{\mathsf{tag}}$ exists, not that it was computed as $\mathsf{Expand}(H_2(\mathbf{s}_2\|\mathsf{event}))$, which would require proving a hash evaluation in zero knowledge.
Linkability is therefore a guarantee on signatures that $\mathsf{Sign}$ produced (Definition~\ref{def:link-honest}); a signer that substitutes an arbitrary short vector emits signatures that $\mathsf{Link}$ reports as unlinked .

Fiat--Shamir with aborts~\cite{ref58} yields $\sigma$.  The relation is proved in two layers that
share one transcript, and this is the structural change that removes the linear factor in $N$.

\emph{Algebraic layer.}  The index $\pi$ enters the five relations only through ${\mathbf{P}}_{1,\pi}$
and ${\mathbf{P}}_{2,\pi}$.  Holding the selector fixed, the relations are the same whichever member
signs, so they collapse into $\mathbf{M}\mathbf{w}=\mathbf{u}$ with
\begin{equation}\label{eq:Mwu}
  \mathbf{M}\mathbf{w}=
  \begin{pmatrix}
    \mathbf{A}\mathbf{s}_1+\mathbf{e}_1\\
    \mathbf{B}\mathbf{s}_2+\mathbf{e}_2\\
    \mathbf{A}_{\mathsf{event}}\mathbf{s}_2+\mathbf{e}_{\mathsf{tag}}\\
    \mathbf{B}^{\top}\mathbf{r}+\mathbf{e}_3\\
    \mathbf{P}_{\mathsf{TA}}^{\top}\mathbf{r}+\mathbf{e}_4+\mathsf{encode}(\mathbf{B}\mathbf{s}_2+\mathbf{e}_2)
  \end{pmatrix}
  =
  \begin{pmatrix}
    \mathbf{P}_1\\[1pt] \mathbf{P}_2\\[1pt] \boldsymbol{\tau}_{\mathsf{link}}\\[1pt] \mathbf{c}_1\\[1pt] \mathbf{c}_2
  \end{pmatrix},
\end{equation}
so the five block rows are \eqref{eq:rel-sig}--\eqref{eq:rel-c2} verbatim, with $\mathbf{w}=(\mathbf{s}_1,\mathbf{e}_1,\mathbf{s}_2,\mathbf{e}_2,\mathbf{e}_{\mathsf{tag}},\mathbf{r},\mathbf{e}_3,\mathbf{e}_4,\boldsymbol{\varrho})$ the witness and, at the relation level, $\mathbf{u}=(\mathbf{P}_1,\mathbf{P}_2,\boldsymbol{\tau}_{\mathsf{link}},\mathbf{c}_1,\mathbf{c}_2)$ the statement together with the commitment row $\hat{\mathbf{u}}=\mathsf{encode}(\mathbf{P}_1,\mathbf{P}_2)+\mathbf{H}\boldsymbol{\varrho}$.
Verification absorbs the pair into the commitment: substituting $\mathbf{P}_1=\hat{\mathbf{u}}_1-\mathbf{H}_1\boldsymbol{\varrho}$ and $\mathbf{P}_2=\hat{\mathbf{u}}_2-\mathbf{H}_2\boldsymbol{\varrho}$, with $\hat{\mathbf{u}}=\hat{\mathbf{u}}_1\oplus\hat{\mathbf{u}}_2$ split along the coordinate blocks of $\mathsf{encode}$, turns rows 1--2 and the $\mathsf{encode}(\mathbf{P}_2)$ term of the last row into equations over $(\mathbf{z},\tilde{\mathbf{u}})$ with $\tilde{\mathbf{u}}=(\hat{\mathbf{u}},\boldsymbol{\tau}_{\mathsf{link}},\mathbf{c}_1,\mathbf{c}_2)$; this absorbed form is what the index-free verifier evaluates, and at the confirmation predicate the opening $\hat{\mathbf{u}}=\mathsf{encode}(\mathbf{P}_{1,i},\mathbf{P}_{2,i})+\mathbf{H}\boldsymbol{\varrho}$ collapses it back onto the rows at $\mathbf{u}_i$.  Both forms cost the six module products above plus the commitment products, every one of them independent of $N$.
Read as a matrix, $\mathbf{M}$ has five block rows and eight block columns; only twelve positions are non-zero and five of those are identity blocks, so evaluating $\mathbf{M}\mathbf{w}$ costs the six module products $\mathbf{A},\mathbf{B},\mathbf{A}_{\mathsf{event}},\mathbf{B}^{\top},\mathbf{P}_{\mathsf{TA}}^{\top},\mathsf{encode}\circ\mathbf{B}$ by one vector each, i.e.\ $6k\ell$ ring multiplications whatever $N$ is.
The last row carries the ciphertext, whose plaintext $\mathsf{encode}(\mathbf{P}_{2,\pi})$ is linear in $(\mathbf{s}_2,\mathbf{e}_2)$ because $\mathsf{encode}$ is linear and $\mathbf{P}_{2,\pi}=\mathbf{B}\mathbf{s}_2+\mathbf{e}_2$; the row therefore needs no auxiliary quotient and no coefficient splitting.
The signer samples a mask $\mathbf{y}$ over the same block structure, publishes $\mathbf{T}=\mathbf{M}\mathbf{y}$, and later answers $\mathbf{z}=\mathbf{y}+c\,\mathbf{w}$.

\emph{Selection layer.}  The signer commits to the one-hot selector $b$ as
$B_{\mathsf{bin}}=G(r_b,\,b,\,\{c_{j,i}\})$ with $c_{j,i}=a_{j,i}(1-2b_{j,i})$, publishes its high half
$\mathsf{cm}_{\mathsf{sel}}=B_{\mathsf{bin,hi}}$, and afterwards answers with $f$, $g$ and $z_b$ as in
Section~\ref{sec:sel}.  The commitment must precede the challenge: otherwise a signer could choose $b$
to fit the $x$ it was given.

\emph{Challenge.}  With both commitments absorbed,
$c=H_3\bigl(\mathsf{inst},\hat{\mathbf{u}},\mathbf{T},\mathsf{cm}_{\mathsf{sel}}\bigr)\in\mathcal{R}_q$
is drawn with small coefficients, and the signer answers both layers.  Rejection sampling makes the
responses witness independent, giving zero knowledge~\cite{DBLP:conf/crypto/LyubashevskyNS21} and, in
the ROM, multi-theorem zero knowledge with simulation extractability
(Section~\ref{sec:zkpok}).  The signer restarts if any response leaves its bound, and outputs $(\sigma,\boldsymbol{\varrho})$ with
$\sigma=(\boldsymbol{\tau}_{\mathsf{link}},\mathbf{c},\hat{\mathbf{u}},\pi_{\mathsf{alg}},\pi_{\mathsf{sel}})$.
Algorithm~1 fixes the order in which tag, ciphertext and the two commitments are
produced, so that the instance is complete before the challenge hash is taken.

\subsection{Verify}
\label{sec:verify-alg}

Verification runs in two tiers, and both are deterministic.
The \emph{public tier} takes $(\mathsf{pp},M,\mathsf{event},R,\sigma)$ and needs no index: it recomputes $\mathbf{A}_{\mathsf{event}}$ and the statement $\mathsf{inst}=(\mathsf{event},R,\hat{\mathbf{u}},\boldsymbol{\tau}_{\mathsf{link}},\mathbf{c}_1,\mathbf{c}_2)$, parses $\pi_{\mathsf{alg}}=(\mathbf{T},\mathbf{z})$ and $\pi_{\mathsf{sel}}=(\{f_{j,i}\},B_{\mathsf{bin,hi}},z_b)$, recomputes $c=H_3(\mathsf{inst},\hat{\mathbf{u}},\mathbf{T},\mathsf{cm}_{\mathsf{sel}})$, and checks
\begin{enumerate}[leftmargin=16px,itemsep=0.15em,topsep=0.2em,parsep=0pt]
\item \emph{selection layer:} the selector identity of Section~\ref{sec:sel} holds on the transported $f$'s, the reconstruction~\eqref{eq:selrec} is consistent, and the position isolated by the leading coefficient of \eqref{eq:selpoly} is unique.
\end{enumerate}
This tier is what every node runs when it decides whether a record is well formed, and by itself it says nothing about which member produced $\sigma$.
The \emph{binding tier} takes the index $i$ in addition, rebuilds $\mathbf{u}_i=(\mathbf{P}_{1,i},\mathbf{P}_{2,i},\boldsymbol{\tau},\mathbf{c}_1,\mathbf{c}_2)$, and checks
\begin{enumerate}[leftmargin=16px,itemsep=0.15em,topsep=0.2em,parsep=0pt,start=2]
\item \emph{algebraic layer:} $\mathbf{M}\mathbf{z}-c\,\mathbf{u}_i=\mathbf{T}$ and $\|\mathbf{z}\|_\infty\le B$, at the cost of $6k\ell$ non-zero ring products of $\mathbf{M}$ ($180$ at Set-III), independent of the ring size.
\end{enumerate}
Check (1) needs the key images $(\mathbf{P}_1,\mathbf{P}_2)$ that the algebraic layer consumes, and this is the only quantity in verification that depends on the signer.
The signature therefore carries a commitment $\hat{\mathbf{u}}=\mathsf{encode}(\mathbf{P}_1,\mathbf{P}_2)+\mathbf{H}\boldsymbol{\varrho}$ to the pair, with $\mathbf{H}$ a public module matrix and $\boldsymbol{\varrho}$ short: the pair enters additively and the mask multiplicatively, so testing a guessed index requires a short preimage under $\mathbf{H}$ rather than a norm check, which is what keeps the commitment hiding for every ring member.
The system then has a sixth row,
\begin{equation}\label{eq:comcheck}
  \mathsf{encode}(\mathbf{P}_1,\mathbf{P}_2)+\mathbf{H}\boldsymbol{\varrho}=\hat{\mathbf{u}},
\end{equation}
and the public statement is $\mathbf{u}=(\hat{\mathbf{u}},\boldsymbol{\tau}_{\mathsf{link}},\mathbf{c}_1,\mathbf{c}_2)$, every component of which is either an M-LWE sample or a binding commitment: nothing in it names a member.
The proof is knowledge of a witness for the relation
{\small
\begin{equation}\label{eq:relation}
\begin{aligned}
\mathcal{R}=\Bigl\{(\mathsf{inst};\,&\mathbf{P}_1,\mathbf{P}_2,\boldsymbol{\varrho}):\ \exists\,\mathbf{s}_1,\mathbf{s}_2,\mathbf{e}_1,\mathbf{e}_2\ \text{with}\\
&\hat{\mathbf{u}}=\mathsf{encode}(\mathbf{P}_1,\mathbf{P}_2)+\mathbf{H}\boldsymbol{\varrho},\quad
\mathbf{P}_1=\mathbf{A}\mathbf{s}_1+\mathbf{e}_1,\\
&\mathbf{P}_2=\mathbf{B}\mathbf{s}_2+\mathbf{e}_2,\\
&\boldsymbol{\tau}_{\mathsf{link}}=\mathbf{A}_{\mathsf{event}}\mathbf{s}_2+\mathbf{e}_{\mathsf{tag}}\Bigr\},
\end{aligned}
\end{equation}}
so the committed pair, its well-formedness, the tag and the ciphertext are one statement: $\mathsf{inst}$ fixes all public inputs, and every variable that names a member appears only in the witness.
Ring membership is deliberately not a line of~\eqref{eq:relation}: a proof of knowledge certifies a well-formed committed pair, and the claim that this pair is a ring entry is carried by the predicates that can see an opening --- $\mathsf{Confirm}$, which checks $\hat{\mathbf{u}}=\mathsf{encode}(\mathbf{P}_{1,i},\mathbf{P}_{2,i})+\mathbf{H}\boldsymbol{\varrho}$ at a named $i$, and $\mathsf{Judge}$, which matches the key that row~\eqref{eq:rel-c2} encrypts against $R$ (Section~\ref{sec:link-alg}).  The two claims coincide for honestly generated signatures, because the signer sets $\hat{\mathbf{u}}$ to its own key and holds $(\boldsymbol{\varrho},\mathbf{s}_{2,\pi})$; the theorems of Section~\ref{sec:security} are stated against these predicates (Definitions~\ref{def:trace} and~\ref{def:unforgeability}).
Rows 3--4 contain only public data and are evaluated in the clear; rows 1--2 and row~\eqref{eq:rel-c2} are evaluated in the absorbed form above, never against a named key.
The selector of (2) certifies a one-hot position declaration for an index it never reveals; tying that declaration to the committed pair is the confirmation predicate's role, not the public tier's.
Well-formedness is thus a property of $\sigma$ alone: no index, no trial, and the check is identical for every verifier, while membership is decided where an opening is present.
The cost is one extra block row and two extra witness blocks, i.e.\ one more module product in verification.
This is where $N$ enters, as $\kappa=\lceil\log_{b_{\mathsf{sel}}} N\rceil$ digits rather than $N$ responses.
Correctness follows from perfect completeness of both layers and from rejection sampling keeping honest responses within the norm bound; Algorithm~2 gives the check sequence.

\noindent\textbf{Neither predicate reveals the signer.}
The two predicates differ in what they decide, not in what they leak.

\begin{proposition}[Anonymity is preserved by verification]
\label{prop:index-hiding}
Let $(\sigma,\boldsymbol{\varrho})\leftarrow\mathsf{Sign}(\mathsf{pp},\mathsf{sk}_\pi,M,\mathsf{event},R)$ for $i_0,i_1\in R$. Then $\mathsf{Verify}$ accepts $\sigma$ for both hypotheses, and there is a simulator that produces $(\boldsymbol{\tau}_{\mathsf{link}},\mathbf{c},\hat{\mathbf{u}},\mathbf{T},\mathbf{z},\pi_{\mathsf{sel}})$ with the same distribution for $\pi=i_0$ and for $\pi=i_1$; $\mathsf{Confirm}$ is deterministic and returns $1$ only for $i=\pi$, which is what the opening establishes.
\end{proposition}

\begin{proof}
$\hat{\mathbf{u}}$ hides $(\mathbf{P}_1,\mathbf{P}_2)$ under the binding commitment, $\mathbf{c}$ hides $\mathbf{P}_2$ under M-LWE and the tag is an M-LWE sample by Lemma~\ref{lem:tag-prf}; the selector hides $\pi$ by Lemma~\ref{lem:sel-unique} and its zero-knowledge argument (Appendix~B-A), and the response is masked by $\mathbf{y}$.
Replacing $\pi$ by another ring member therefore changes no element of $\sigma$ in distribution, so no verifier running either predicate distinguishes the two.
\end{proof}

Because $\mathsf{Verify}$ is index-free and $\hat{\mathbf{u}}$ is hiding, the protocol exposes no per-index accept/reject oracle to anyone: an index test requires the opening $\boldsymbol{\varrho}$, which $\sigma$ does not carry and which no party but the signer can compute, so the $N$ trial runs that would recover $\pi$ do not exist.

\subsection{Link, Open and Judge}
\label{sec:link-alg}
\label{sec:trace-alg}

$\mathsf{Link}$ decides whether two signatures \emph{for the same event} share a signer: for $\sigma_1=(\boldsymbol{\tau}_{\mathsf{link},1},\cdot,\cdot)$ and $\sigma_2=(\boldsymbol{\tau}_{\mathsf{link},2},\cdot,\cdot)$ it returns $\mathsf{linked}$ iff $\boldsymbol{\tau}_{\mathsf{link},1}=\boldsymbol{\tau}_{\mathsf{link},2}$.
A linked pair can be flagged on-chain as a verifiable \textbf{dishonesty tag} naming no identity.
Two signatures produced by $\mathsf{Sign}$ for one user on one event always link (Definition~\ref{def:link-honest}, condition~(1)), since $\mathbf{e}_{\mathsf{tag}}$ and $\mathbf{A}_{\mathsf{event}}$ are determined by $(\mathbf{s}_2,\mathsf{event})$, while pseudorandomness keeps tags of different events uninformative; because~\eqref{eq:rel-tag} forces the noise to be short rather than the expansion of $H_2$, a client that departs from $\mathsf{Sign}$ can make its tags differ, which is the case the dispute rule of Section~\ref{sec:ledger} is for.

$\mathsf{Open}$, restricted to the TA, decrypts $\mathbf{v}=\mathbf{c}_2-\mathbf{S}_{\mathsf{TA}}^{\top}\mathbf{c}_1=\mathsf{encode}(\mathbf{P}_{2,\pi})+\mathbf{E}_{\mathsf{TA}}^{\top}\mathbf{r}+\mathbf{e}_4-\mathbf{S}_{\mathsf{TA}}^{\top}\mathbf{e}_3$ with the last three terms short (Appendix~B-G) and recovers $\tilde{\mathbf{P}}_2$ by $\mathsf{decode}$ and rounding.
It also returns $\pi_{\mathsf{dec}}$, a proof, with witness $(\mathbf{S}_{\mathsf{TA}},\mathbf{E}_{\mathsf{TA}},\mathbf{e}_{\mathsf{dec}})$, of the relation $\mathbf{P}_{\mathsf{TA}}=\mathbf{B}\mathbf{S}_{\mathsf{TA}}+\mathbf{E}_{\mathsf{TA}}\ \land\ \mathbf{c}_2-\mathbf{S}_{\mathsf{TA}}^{\top}\mathbf{c}_1=\mathsf{encode}(\tilde{\mathbf{P}}_2)+\mathbf{e}_{\mathsf{dec}}\ \land\ \|\mathbf{e}_{\mathsf{dec}}\|_\infty\le\beta_{\mathsf{trace}}$, which states that the ciphertext opens to the key returned; the relation is linear with a short witness, so the sigma protocol of Section~\ref{sec:zkpok} proves it without revealing $\mathbf{S}_{\mathsf{TA}}$, and a proof for a key other than the plaintext would place two ring keys within $2\beta_{\mathsf{trace}}$, which Proposition~\ref{prop:trace-uniqueness} excludes.
$\mathsf{Judge}$ takes $(\sigma,\tilde{\mathbf{P}}_2,\pi_{\mathsf{dec}})$, checks $\mathsf{Verify}(\sigma)=1$ and $\pi_{\mathsf{dec}}$, locates the unique $\mathsf{pk}_i\in R$ with $\|\tilde{\mathbf{P}}_2-{\mathbf{P}}_{2,i}\|_\infty\le\beta_{\mathsf{trace}}$, and returns $\mathsf{convict}$ iff all three hold.
An opening is therefore a claim the contract rechecks rather than a decryption a verifier must take on trust: a key outside the ciphertext fails the proof, a key outside the ring fails the match, and the authority can withhold $\tilde{\mathbf{P}}_2$ but cannot make $\mathsf{Judge}$ convict a member who did not sign.
Linking is one tag comparison, and opening is one decryption plus a bounded search that runs off-chain.

\noindent\textbf{Across events.}
Tags are uninformative across events by design, so a cross-event conflict is not decided by $\mathsf{Link}$.
The contract then opens both signatures and compares the recovered keys: $\mathsf{JudgePair}(\sigma_1,\sigma_2,\tilde{\mathbf{P}}_{2,1},\pi_{\mathsf{dec},1},\tilde{\mathbf{P}}_{2,2},\pi_{\mathsf{dec},2})$ returns $\mathsf{same}$ when both openings land on one ring member, $\mathsf{different}$ when they land on two, and $\bot$ when either key matches no member of its ring or either decryption proof fails.
$\mathsf{JudgePair}$ is a rule of the contract rather than an algorithm of the primitive of Definition~\ref{def:syntax}: it applies $\mathsf{Judge}$ twice and compares the two keys, so the syntax keeps eight algorithms and the security games never call it.
The three outcomes are what the ledger rule of Section~\ref{sec:ledger} distinguishes: one member acting twice, or two members sharing one authorisation.

\subsection{The selection layer}\label{sec:sel}

Write each index $i\in[0,N)$ in base $b_{\mathsf{sel}}$ as $i=(i_0,\dots,i_{\kappa-1})$ with $N=b_{\mathsf{sel}}^{\kappa}$, where $b_{\mathsf{sel}}$ is the selector radix, kept apart from the norm bound $\beta$; the index ranges over the concatenated keys $Q_i=(\mathbf{P}_{1,i},\mathbf{P}_{2,i})$, so one selector fixes both components.
For digit $j$ the prover holds a one-hot vector $b_{j,\cdot}\in\{0,1\}^{b_{\mathsf{sel}}}$ with
$\sum_{i}b_{j,i}=1$, together with a mask $a_{j,i}$ drawn from a discrete Gaussian.  Writing $x$ for the
challenge $c\in\mathcal{R}_q$ of Section~\ref{sec:sign-alg} (the ciphertext is $\mathbf{c}$), define
\begin{equation}\label{eq:selvals}
  f_{j,i}=a_{j,i}+x\,b_{j,i},\
  c_{j,i}=a_{j,i}(1-2b_{j,i}),\
  d_{j,i}=-a_{j,i}^{2}.
\end{equation}
Then
\begin{equation}\label{eq:selpoly}
  \prod_{j=0}^{\kappa-1} f_{j,i_j}=\delta_{i,\pi}\,x^{\kappa}+\sum_{m=0}^{\kappa-1}C_m x^{m},
\end{equation}
so the coefficient of $x^{\kappa}$ is one exactly for the signer's column and zero elsewhere: the
selector isolates $\pi$ without naming it.

The proof carries $f$, the quadratic values $g_{j,i}=f_{j,i}(x-f_{j,i})$, the masking response
$z_b=r_a+xr_b$, and the high half $B_{\mathsf{bin,hi}}$ of $B_{\mathsf{bin}}=G(r_b,b,\{c_{j,i}\})$.  The mask
$a$ is never transmitted.  The verifier reconstructs
\begin{equation}\label{eq:selrec}
  \hat A=G(z_b,f,g)-2^{K_B}x\,B_{\mathsf{bin,hi}}=A_{\mathsf{bin}}+x\,B_{\mathsf{lo}},
\end{equation}
with $A_{\mathsf{bin}}=G(r_a,a,d)$ and $B_{\mathsf{bin}}=2^{K_B}B_{\mathsf{bin,hi}}+B_{\mathsf{lo}}$ the
centred decomposition.

\begin{lemma}[Binarity is exactly what the identity tests]\label{lem:bin}
For every $j$ and $i$, $d_{j,i}+x\,c_{j,i}=g_{j,i}$ if and only if $b_{j,i}^{2}=b_{j,i}$.
\end{lemma}
\begin{proof}
By~\eqref{eq:selvals}, $d_{j,i}+xc_{j,i}=-a_{j,i}^{2}+xa_{j,i}(1-2b_{j,i})$, whereas
$g_{j,i}=(a_{j,i}+xb_{j,i})(x-a_{j,i}-xb_{j,i})=-a_{j,i}^{2}+a_{j,i}x-2a_{j,i}xb_{j,i}+x^{2}(b_{j,i}-b_{j,i}^{2})$.
The first three terms coincide, so the two expressions agree iff $x^{2}(b_{j,i}-b_{j,i}^{2})=0$.
Because $x\in\mathcal{C}$ is a unit (Definition~\ref{def:unitsep}), $x^{2}$ is a unit as well, so this is exactly $b_{j,i}=b_{j,i}^{2}$.
\end{proof}

Lemma~\ref{lem:bin} is why the selector needs no separate range proof: a non-binary digit breaks~\eqref{eq:selrec} by $-2x^{2}$, which the verifier detects without learning $a$.

\begin{lemma}[The selector pins at most one index]\label{lem:sel-unique}
Suppose $q\equiv1\pmod{2d}$. Let $b_{j,i}=f_{j,i}-a_{j,i}$ satisfy $b_{j,i}^{2}=b_{j,i}$ and $\sum_{i}b_{j,i}=1$ in $\mathcal{R}_q$ for every digit $j$. Then at most one index has the coefficient of $x^{\kappa}$ in \eqref{eq:selpoly} equal to one.
\end{lemma}

\begin{proof}
Since $q\equiv1\pmod{2d}$, the polynomial $X^{d}+1$ splits into distinct linear factors over $\mathbb{F}_q$ and $\mathcal{R}_q\cong\mathbb{F}_q^{d}$ by the Chinese remainder theorem. Evaluate at a root $\zeta$: the identity $b_{j,i}^{2}=b_{j,i}$ gives $b_{j,i}(\zeta)\in\{0,1\}$, and $\sum_{i}b_{j,i}=1$ gives exactly one digit equal to one in each row $j$, so row $j$ selects one index $i_j(\zeta)$. The coefficient of $x^{\kappa}$ at index $i$, evaluated at $\zeta$, is $\prod_{j}b_{j,i_j}(\zeta)$, which equals one exactly when every row selects $i$ at $\zeta$. If two distinct indices $\pi\neq\pi'$ both had coefficient one, then at every root every row would select both $\pi$ and $\pi'$, contradicting the one-hot selection at that root. Hence at most one index qualifies.
\end{proof}

Hence at most one index satisfies the identity of Lemma~\ref{lem:sel-unique}, and completeness supplies one, so the public tier cannot be satisfied by two members at once, and the transcript reveals $\pi$ only through a product identity whose coefficients are functions of the public ring: a simulator that samples $a$ uniformly and fixes any $\pi$ reproduces the distribution (Appendix~B-A).  One digit per row follows by subtraction: $a_{j,0}:=-\sum_{i=1}^{b_{\mathsf{sel}}-1}a_{j,i}$ and $f_{j,0}:=x-\sum_{i=1}^{b_{\mathsf{sel}}-1}f_{j,i}$ give $\sum_{i}b_{j,i}=1$ for free, so only $b_{\mathsf{sel}}-1$ digits per row are committed.

\section{Security Analysis}
\label{sec:security}

Fix a PPT adversary $\mathcal{A}$ with $q_s$ signing and $q_h$ random-oracle queries, and write $\mathsf{Adv}^{\mathsf{X}}_{\mathcal{A}}(\lambda)=\Pr[\mathsf{G}^{\mathsf{X}}_{\mathcal{A}}\Rightarrow1]$ for the game named in Definitions~\ref{def:anon}--\ref{def:unforgeability}.
Each theorem is proved by a chain $\mathsf{G}_0,\dots,\mathsf{G}_k$ from the definitional game to an ideal one whose answer is independent of the hidden bit, or in which a witness is known; every transition consumes one assumption and its loss is stated, so each bound is the sum of its hops.
Full transcripts and auxiliary lemmas are in Appendix~B.

\subsection{The two extraction steps}\label{sec:proof-arch}

Every reduction forks once and extracts in two steps.
(F1)~\emph{Forking} gives two accepting transcripts with equal first moves and distinct challenges with probability at least $\varepsilon/q_h-1/|\mathcal{C}|$, a loss we write $\mathsf{lf}(\varepsilon)$ and the only place $q_h$ enters.
(F2)~\emph{Extraction} then gives $\mathbf{M}(\mathbf{z}-\mathbf{z}')=(c-c')\,\mathbf{u}$. Reading $\mathbf{u}$ divides by $c-c'$, a unit whose inverse need not preserve norms, so the norm bound is carried as the hypothesis $\|\mathbf{w}\|_\infty\le2\beta$; under it, two distinct witnesses for one $\mathbf{u}$ differ by a nonzero kernel vector of norm at most $2\beta$, an $\mathsf{M\text{-}SIS}_{k,\ell,q,2\beta}$ solution, and the selection layer returns the position from the same transcripts.

Three transcripts remove the division: they give $\mathbf{M}\mathbf{w}=\mathbf{0}$ with a bound that grows with $d$, which Appendix~B-B records.

\noindent\textbf{The challenge space.}
$\mathcal{C}\subset\mathcal{R}_q$ is a finite set of polynomials of degree below $d$ with $\|c\|_\infty\le B_c$, sampled by $H_3$.

\begin{definition}[Admissible challenge space]
\label{def:unitsep}
$\mathcal{C}$ is \emph{admissible} if every $c\in\mathcal{C}$ and every difference $c-c'$ of distinct elements is a unit of $\mathcal{R}_q$.
\end{definition}

Admissibility does not follow from the coefficient bound: $\mathcal{R}_q=\mathbb{F}_q[X]/(X^d+1)$ is a product of fields, and $h$ is a unit exactly when $\gcd(h,X^d+1)=1$ over $\mathbb{F}_q$, which fails only when $h$ vanishes at one of the $d$ roots of $X^d+1$.
The sampler enforces it by rejection: $H_3$ draws a weight-$\omega$ polynomial with coefficients in $\{0,\pm1\}$, the prover redraws when $\gcd(c,X^d+1)\neq1$, and the extractor redraws its second and third challenges until $c-c'$ is a unit, a test it can run because it chooses them.
A draw fails with probability at most $d/q$ by a union bound over the roots, so the expected number of draws is $1/(1-d/q)$ at one gcd over $\mathcal{R}_q$ per draw, and $\varepsilon_{\mathsf{div}}=0$ for this sampler; the bounds below keep $2\varepsilon_{\mathsf{div}}$ explicit so that an unconditional fork, which fails with probability at most $2d/q$, is covered by the same expression.
Admissibility is consumed twice: (F2) divides by $c-c'$ to read $\mathbf{u}$, and Lemma~\ref{lem:bin} divides by $x^2$ to cancel the mask.
The selector also needs three transcripts, i.e.\ two forks from one first move, so its loss is $2\,\mathsf{lf}(\varepsilon)+2\varepsilon_{\mathsf{div}}$, and the bounds below carry the additive term.

\begin{lemma}[Selector extraction]\label{lem:sel-extract}
Given three accepting transcripts with pairwise invertible challenge differences, obtained by two forks from a common first move, the extracted
selector values satisfy $f_{j,i}=a_{j,i}+x\,b_{j,i}$ with $b_{j,i}^{2}=b_{j,i}$ and
$\sum_{i}b_{j,i}=1$ in $\mathcal{R}_q$ for every digit $j$; consequently, by
Lemma~\ref{lem:sel-unique}, the leading coefficient of \eqref{eq:selpoly} is one at exactly one index $\pi^{\ast}\in[N]$.
\end{lemma}
\begin{proof}[Proof sketch]
Write the relation of Lemma~\ref{lem:bin} at the three challenges $x,x',x''$ and eliminate the
quadratic term: the residual is linear in $x$ with coefficient $b_{j,i}-b_{j,i}^{2}$, which must
vanish because $x-x'$, $x-x''$ and $x'-x''$ are invertible; hence $b_{j,i}^{2}=b_{j,i}$.  The row sum
follows from $f_{j,0}=x-\sum_{i\ge1}f_{j,i}$ and $a_{j,0}=-\sum_{i\ge1}a_{j,i}$, which force
$\sum_{i}b_{j,i}=1$, and substituting into~\eqref{eq:selpoly} gives the claim.
\end{proof}

The steps compose: the challenge is drawn after both commitments are absorbed
(Section~\ref{sec:sign-alg}), so the selector cannot follow $x$, and $\mathbf{u}$ is fixed by the selector before the algebraic response exists, so the extractor returns a single $\pi^{\ast}$ and a single witness.
A forgery therefore yields an M-SIS solution at~(F2) or two distinct one-hot selectors, the second excluded by uniqueness of $\pi^{\ast}$.

\subsection{Anonymity}

\begin{theorem}[Anonymity]\label{thm:anon}
If $\mathsf{M\text{-}LWE}_{k,\ell,q,\eta}$ is hard and the NIZKPoK of Section~\ref{sec:zkpok} is multi-theorem zero knowledge in the ROM, \textsf{PLATRS} satisfies CPA anonymity (Definition~\ref{def:anon}), and for every PPT $\mathcal{A}$
\[
 \mathsf{Adv}^{\mathsf{ANON}}_{\mathcal{A}}(\lambda)\le 2\,\mathsf{Adv}^{\mathsf{M\text{-}LWE}}+(q_s+1)\,\varepsilon_{\mathsf{zk}}+\tfrac{(q_s+q_h)^2}{|\mathcal{C}|},
\]
where $\mathcal{C}$ is the challenge space and $\varepsilon_{\mathsf{zk}}$ the simulation error of the sigma protocol, statistical at our parameters (Section~\ref{sec:zkpok}) rather than a hardness assumption.
\end{theorem}

\begin{proof}[Game sequence]
$\mathsf{G}_0$ is the game of Definition~\ref{def:anon}. $\mathsf{G}_1$ replaces both proofs by the NIZKPoK simulator, which fails with probability at most $(q_s+q_h)^2/|\mathcal{C}|$, so $|\Pr[\mathsf{G}_0]-\Pr[\mathsf{G}_1]|\le(q_s+1)\mathsf{Adv}^{\mathsf{ZK}}+(q_s+q_h)^2/|\mathcal{C}|$; $\mathsf{G}_2$ replaces $\boldsymbol{\tau}^{\ast}$ by a uniform $\mathbf{u}^{\ast}$, an M-LWE hop by Lemma~\ref{lem:tag-prf}; and $\mathsf{G}_3$ replaces $\mathbf{c}^{\ast}$ by an encryption of $\mathbf{0}$, an IND-CPA hop that reduces to M-LWE. In $\mathsf{G}_3$ the challenge is independent of $b$, so $\Pr[b'=b]=1/2$ and the bound follows.
\end{proof}

\subsection{Linkability}

\begin{theorem}[Linkability for honestly generated signatures]\label{thm:link}
If $\mathsf{M\text{-}SIS}_{k,\ell,q,2\beta}$ is hard and the NIZKPoK is simulation extractable, \textsf{PLATRS} satisfies $\mathsf{Link}_{\mathsf{hon}}$ (Definition~\ref{def:link-honest}), and
$\mathsf{Adv}^{\mathsf{LINK}}_{\mathcal{A}}(\lambda)\le q_h\big(\mathsf{Adv}^{\mathsf{M\text{-}SIS}}(\lambda)+\mathsf{Adv}^{\mathsf{M\text{-}LWE}}(\lambda)+1/|\mathcal{C}|\big)+2\varepsilon_{\mathsf{div}}$.
\end{theorem}

\begin{proof}[Game sequence]
$\mathsf{G}_0$ is the game of Definition~\ref{def:link-honest}. A winning output is a pair $(\sigma_1,\sigma_2)$ that either links two runs of one honest signer or fails condition~(2).

$\mathsf{G}_1$ replaces both signatures by their extraction: by (F1) this succeeds with probability $\mathsf{lf}(\varepsilon)$, and by (F2) it yields $\boldsymbol{\tau}^{(j)}=\mathbf{A}_{\mathsf{event}}\mathbf{s}_2^{(j)}+\mathbf{e}_{\mathsf{tag}}^{(j)}$ with $\|\mathbf{e}_{\mathsf{tag}}^{(j)}\|_\infty\le\eta$ for $j=1,2$.

If $\mathbf{s}_2^{(1)}\neq\mathbf{s}_2^{(2)}$ while $\boldsymbol{\tau}^{(1)}=\boldsymbol{\tau}^{(2)}$, then $\mathbf{A}_{\mathsf{event}}(\mathbf{s}_2^{(1)}-\mathbf{s}_2^{(2)})=\mathbf{e}_{\mathsf{tag}}^{(2)}-\mathbf{e}_{\mathsf{tag}}^{(1)}$ is a nonzero vector of norm at most $2\eta$ in the image of $\mathbf{A}_{\mathsf{event}}$, which is an $\mathsf{M\text{-}SIS}$ solution for the matrix $\mathbf{A}_{\mathsf{event}}$ at norm $2\beta$; this is the only hop that consumes an assumption.

If $\mathbf{s}_2^{(1)}=\mathbf{s}_2^{(2)}$ then both signatures name one key, and condition~(2) holds because $\mathsf{Open}$ returns that key (Lemma~\ref{lem:trace-exact} and Theorem~\ref{thm:trace}); condition~(1) follows from determinism: $\mathbf{A}_{\mathsf{event}}$ is a function of $\mathsf{event}$ and $\mathbf{e}_{\mathsf{tag}}$ a function of $(\mathbf{s}_2,\mathsf{event})$, so the two tags are equal.
\end{proof}

Relation~\eqref{eq:rel-tag} constrains a short vector rather than the output of $H_2$, so a client that departs from $\mathsf{Sign}$ can produce a second tag.
Detection for that signature falls to the ledger's dispute rule (Section~\ref{sec:ledger}), which opens both signatures and compares the recovered keys.

\subsection{Traceability}

\begin{theorem}[Traceability]\label{thm:trace}
If $\mathsf{M\text{-}SIS}_{k,\ell,q,\max\{2\beta,\,2\beta_{\mathsf{trace}}\}}$ and $\mathsf{M\text{-}LWE}_{k,\ell,q,\eta}$ are hard and the NIZKPoK is simulation extractable, \textsf{PLATRS} is traceable (Definition~\ref{def:trace}), and
$\mathsf{Adv}^{\mathsf{TRACE}}_{\mathcal{A}}(\lambda)\le q_h\big(\mathsf{Adv}^{\mathsf{M\text{-}SIS}}(\lambda)+\mathsf{Adv}^{\mathsf{M\text{-}LWE}}(\lambda)+1/|\mathcal{C}|\big)+2\varepsilon_{\mathsf{div}}+\varepsilon_{\mathsf{dec}}$, where $\varepsilon_{\mathsf{dec}}$ is the decryption-failure probability of Proposition~\ref{prop:trace-bound}.
\end{theorem}

\begin{proof}[Game sequence]
$\mathsf{G}_0$ is the game of Definition~\ref{def:trace}. $\mathsf{G}_1$ extracts both signatures by (F1)--(F2), costing $q_h$; the extracted witness of~\eqref{eq:relation} gives a well-formed committed pair $(\mathbf{P}_1,\mathbf{P}_2)$ with $\mathbf{P}_1=\mathbf{A}\mathbf{s}_1+\mathbf{e}_1$ and $\mathbf{P}_2=\mathbf{B}\mathbf{s}_2+\mathbf{e}_2$, and row~\eqref{eq:rel-c2} places $\mathsf{encode}(\mathbf{P}_2)$ inside $\mathbf{c}$: the ciphertext encrypts the \emph{witness} key.  Decryption recovers it up to $\beta_{\mathsf{trace}}$ except with the probability $\varepsilon_{\mathsf{dec}}$ of Proposition~\ref{prop:trace-bound}, so a conviction names $i$ with $\|\mathbf{P}_2-\mathbf{P}_{2,i}\|_\infty\le2\beta_{\mathsf{trace}}$, and Proposition~\ref{prop:trace-uniqueness} makes the match in $R$ unique.  $\mathsf{G}_2$ reads the key equations at $i$: either $\mathbf{s}_2=\mathbf{s}_{2,i}$, which for an honest, uncorrupted $i$ is key recovery, excluded by Lemma~\ref{lem:key-hiding} (M-LWE), or $[\mathbf{B}\mid\mathbf{I}_k\mid-\mathbf{I}_k][\mathbf{s}_2-\mathbf{s}_{2,i};\,\mathbf{e}_2-\mathbf{e}_{2,i};\,\mathbf{P}_2-\mathbf{P}_{2,i}]^{\top}=\mathbf{0}$ is a nonzero kernel relation with blocks of norm at most $2\beta,2\beta,2\beta_{\mathsf{trace}}$, an M-SIS solution at $\max\{2\beta,2\beta_{\mathsf{trace}}\}$.  A fabricated opening is excluded by the soundness of $\pi_{\mathsf{dec}}$, and framing an honest single signer needs this M-SIS event or key recovery. The bound is then $(q_h(\mathsf{Adv}^{\mathsf{M\text{-}SIS}}+\mathsf{Adv}^{\mathsf{M\text{-}LWE}}+1/|\mathcal{C}|)+2\varepsilon_{\mathsf{div}}+\varepsilon_{\mathsf{dec}})$.
\end{proof}

\subsection{Unforgeability}

\begin{theorem}[Unforgeability]\label{thm:unforgeability}
If $\mathsf{M\text{-}SIS}$ is hard at norms $2\beta$ and $2\eta$ and the NIZKPoK is simulation extractable, \textsf{PLATRS} is unforgeable in the sense of Definition~\ref{def:unforgeability}, and
$\mathsf{Adv}^{\mathsf{EUF}}_{\mathcal{A}}(\lambda)\le q_h\big(\mathsf{Adv}^{\mathsf{M\text{-}SIS}_{2\beta}}(\lambda)+\mathsf{Adv}^{\mathsf{M\text{-}SIS}_{2\eta}}(\lambda)+1/|\mathcal{C}|\big)+2\varepsilon_{\mathsf{div}}$.
\end{theorem}

\begin{proof}[Game sequence]
$\mathsf{G}_0$ is the game of Definition~\ref{def:unforgeability}: $\mathcal{A}$ names an index $i\in R$ and returns $(M,\mathsf{event},R,i,\sigma)$ accepted by $\mathsf{Verify}(\cdot,\cdot,\cdot,R,\cdot)$ and by $\mathsf{Confirm}(\cdot,\cdot,\cdot,R,i,\cdot)$ for a pair $(M,\mathsf{event},R)$ it did not query.

$\mathsf{G}_1$ extracts $\sigma$ by (F1)--(F2), costing a factor $q_h$, and obtains the witness of~\eqref{eq:relation}: a committed pair with $\hat{\mathbf{u}}=\mathsf{encode}(\mathbf{P}_1,\mathbf{P}_2)+\mathbf{H}\boldsymbol{\varrho}'$, $\mathbf{P}_1=\mathbf{A}\mathbf{s}_1+\mathbf{e}_1$ and $\mathbf{P}_2=\mathbf{B}\mathbf{s}_2+\mathbf{e}_2$ at bounds $\beta$.  The winning condition includes $\mathsf{Confirm}(i,\sigma,\boldsymbol{\varrho})=1$, i.e.\ $\hat{\mathbf{u}}=\mathsf{encode}(\mathbf{P}_{1,i},\mathbf{P}_{2,i})+\mathbf{H}\boldsymbol{\varrho}$ for the named $i$.  Subtracting the two openings leaves $\mathsf{encode}(\mathbf{P}_{1,i},\mathbf{P}_{2,i})-\mathsf{encode}(\mathbf{P}_1,\mathbf{P}_2)=\mathbf{H}(\boldsymbol{\varrho}'-\boldsymbol{\varrho})$: either the pairs coincide, and the witness then contains a short secret for $\mathsf{pk}_i$, or the difference is a nonzero $\mathbf{H}(\boldsymbol{\varrho}'-\boldsymbol{\varrho})$ of norm at most $2\eta$, an $\mathsf{M\text{-}SIS}$ solution for $\mathbf{H}$.  The selector places the declared $\pi$ and is not needed for this bridge.

$\mathsf{G}_2$ compares the extracted witness with the honest one for $i$, which exists because $i$ is honest and $R$ contains only honest users; in the coinciding case of $\mathsf{G}_1$ the extracted witness is a second short solution of the honest key equations, commitment row included.  If the two witnesses differ, their difference is a nonzero kernel vector of the extended matrix ($\mathbf{M}$ with the $\mathbf{H}$ columns of row~\eqref{eq:comcheck}), of norm at most $2\beta$ in the key rows and $2\eta$ through the commitment columns, an M-SIS solution at $\max\{2\beta,2\eta\}$; if they agree, then $\sigma$ is the honest signature on $(M,\mathsf{event},R)$ and the signing oracle was not queried on that triple, so $\mathcal{A}$ produced a fresh challenge collision, which happens with probability at most $q_h/|\mathcal{C}|$ because $H_3$ is a random oracle.
\end{proof}

\subsection{Correspondence between definitions and proofs}

Each property is proved from its own game: anonymity (Definition~3) over M-LWE with the zero-knowledge simulator, linkability (Definition~4) and authority-only opening (Definition~5) over M-SIS with M-LWE, unforgeability (Definition~6) over M-SIS at norms $2\beta$ and $2\eta$ through the commitment bridge. Every bound carries the forking loss $q_h$ of Proposition~\ref{prop:sim-ext} and the defect $\varepsilon_{\mathsf{div}}$ of Definition~\ref{def:unitsep}, so the theorems and their proofs in Appendix~B are the same expressions.

Appendix~\ref{app:cpa} shows why public per-event linking rules out CCA anonymity, which is the ceiling we analyse.
Every reduction loses $q_h$ and nothing in $N$: the extracted object is the fixed-dimension witness of~\eqref{eq:Mwu}, whereas a flat instantiation needs $N$ independent extractions.

\section{Performance Evaluation}
\label{sec:performance}

\noindent\textbf{Setup.}
\textsf{PLATRS} is implemented in C11 over $\mathcal{R}_q=\mathbb{Z}_q[X]/(X^d+1)$ with an iterative twist-then-NTT kernel on process-wide twiddle tables that the worker threads share read-only.
Experiments ran on an 11th~Gen Intel Core i7-11700 (16 cores, 2.50~GHz), 32~GB of memory, Ubuntu~20.04, all measurements using Set-III ($d=256$, $k=5$, $\ell=6$).
Each baseline is linked against our driver and driven through its own arithmetic, excluding key generation; every number is single-threaded, \textsf{PLATRS} entries are medians of twenty runs in fresh processes and baseline entries single runs.
Code, drivers, plotting scripts and the raw CSVs behind every table are released at \url{https://github.com/JackAugust/PLATRS.git}.

\begin{table}[t]
\centering
\caption{Cost of \textsf{PLATRS} versus ring size $N$.}
\label{tab:platrs}

\begin{tabular*}{\columnwidth}{@{\extracolsep{\fill}}rccccc@{}}
\toprule
$N$ & size & sign & verify & link & open \\
\midrule
4  & 83.3 & 7.999 & 3.495 & 0.0006 & 0.478 \\
8  & 86.3 & 8.219 & 3.537 & 0.0006 & 0.478 \\
16 & 89.3 & 8.458 & 3.595 & 0.0006 & 0.477 \\
32 & 92.3 & 8.687 & 3.650 & 0.0006 & 0.478 \\
64 & 95.3 & 8.971 & 3.698 & 0.0007 & 0.481 \\
128 & 98.3 & 9.178 & 3.749 & 0.0007 & 0.479 \\
256 & 101.3 & 9.425 & 3.808 & 0.0007 & 0.480 \\
512 & 104.3 & 9.670 & 3.865 & 0.0008 & 0.480 \\
1024 & 107.3 & 9.886 & 3.919 & 0.0007 & 0.480 \\
\bottomrule
\end{tabular*}
\par\vspace{2pt}
{\raggedright\footnotesize Sizes in KiB, times in ms; Set-III ($d=256$, $q\approx2^{23}$, $k=5$, $\ell=6$); medians of 20 runs in fresh processes after three discarded warm-up runs, single thread.\par}
\end{table}

\noindent\textbf{Cost of the scheme.}
Table~\ref{tab:platrs} reports \textsf{PLATRS} for $N\in\{4,\dots,1024\}$.
The statement $\mathbf{u}$ has $3k+2\ell=27$ entries and the witness $6(k+\ell)=66$, both fixed, and the ring enters only through $\kappa=\lceil\log_2 N\rceil$ selector digits; signing and verification are therefore flat in $N$, while the length grows by one digit block per doubling of the ring.
Across the sweep signing stays between $8.0$ and $9.9$~ms and verification between $3.5$ and $3.9$~ms, while the length rises from $83.3$ to $107.3$~KiB, that is $+3{,}072$ bytes per digit.
At $N=1024$ it signs in 9.886~ms and verifies in 3.919~ms (101.2\ signatures per second), with linking at 0.7~$\mu$s and opening at 480~$\mu$s; both act on the tag and the ciphertext, whose element counts are fixed by $(k,\ell)$ and do not grow with the ring.
Linking is the one operation where \textsf{PLATRS} is the slowest of the measured schemes: the tag is $k$ ring elements, so a comparison touches $kd=1{,}280$ coefficients ($3.84$~kB) against one element or a short nullifier elsewhere; at 0.7~$\mu$s it is still about $5{,}600\times$ below one verification.
Signature length is 107.3~KiB at $N=1024$ (109\,824\ bytes), which is $103+4\kappa$ ring elements of $dc_b=768$ bytes at $c_b=3$ bytes per coefficient.
The law holds outside the table: $107$, $115$, $139$ and $143$ elements at $N=2,8,512,1024$ ($\kappa=1,3,9,10$).
The growth is logarithmic rather than linear: a flat OR would need one branch per member and is 39.7$\times$ longer at $N=64$ (Appendix~C, last row).

\noindent\textbf{Cost of the remaining operations.}
Key generation costs 0.855~ms per key at $N=64$ and 0.853~ms at $N=512$, independent of the ring; a 512-member ring is populated in 0.44~s.
$\mathsf{Open}$ is one M-LWE decryption of $(\mathbf{c}_1,\mathbf{c}_2)$ and costs 0.480~ms at either $N$, because the ciphertext dimensions depend on $(k,\ell)$ only.
$\mathsf{Judge}$ re-runs $\mathsf{Confirm}$ at the recovered index after the $O(N)$ scan that locates it: 4.386~ms at $N=1024$ and 4.335~ms at $N=512$, so adjudication including the opening totals 4.867~ms at $N=1024$ and stays between 4.447\ and 4.867~ms over $N\in\{4,\dots,1024\}$.
The scan stays below a millisecond at $N=512$, so adjudication is dominated by the re-verification.
Memory is set by the ring rather than by the signature: the implementation allocates each public key at the compile-time maxima ($d=512$, $k=8$), 64~KiB per key, so a 512-member ring occupies 32~MiB, and peak RSS over a process that generates and holds all $N$ key pairs reaches 76.3~MB at $N=512$.

\begin{table*}[t]
\centering
\caption{Measured cost at $N=1024$ under one harness.}
\label{tab:compare}

\setlength{\tabcolsep}{5pt}
\begin{tabular}{@{}lcccccc@{}}
\toprule
Scheme & sign (ms) & verify (ms) & link (ms) & trace (ms) & size (KiB) & \multicolumn{1}{c}{size$_s$/size$_{\text{PLATRS}}$} \\
\midrule
PLATRS & 9.886 & 3.919 & 0.0007 & 4.867 & 107.3 & 1.00 \\
YLGT23~\cite{YLGT23} & 43.34 & 40.79 & 0.0004 & -- & 168.8 & 1.57 \\
CXGG$^+$24~\cite{FS-LLRS} & 2805 & 386.7 & 0.0003 & -- & 5124 & 47.8 \\
DZWW$^+$24~\cite{DZWW+24} & 4401 & 4395.4 & 0.0000 & -- & 48.0 & 0.45 \\
FLLL$^+$21 (model)~\cite{FLLL+21} & 31514 & 28340.9 & n/m & 130687 & 14257 & 132.9 \\
JESS$^+$26~\cite{jagganath2026lotrs} & 700581 & 40975.9 & -- & -- & 194.0 & 1.81 \\
\bottomrule
\end{tabular}
\par\vspace{3pt}
\parbox{\textwidth}{\footnotesize
All rows are single-threaded; every scheme is rebuilt by the same compiler flags and driven by the same harness on one machine. The \textsf{PLATRS} verify column reports verification at a known index. The \textsf{PLATRS} entries are medians of 20 runs, \textsf{YLGT23}, \textsf{CXGG}$^+$24 and \textsf{DZWW}$^+$24 entries medians of five, and JESS$^+$26 carries the mean over its harness's signing seeds. The FLLL$^+$21 row is an operation-count model of its published parameters, not a runnable proof, and it is excluded from the ranking. A dash means the scheme does not provide the operation; n/m means not measured.}
\end{table*}

\subsection{Position in the Field}
\label{sec:perf-compare}

\noindent\textbf{Where it stands.}
Ranked among the five measured schemes it is 1\ of 5\ on signing, 1\ on verification and 2\ on length (mean rank 1.33), and inside the accountability class it is both the shortest and the fastest: the only shorter entry (DZWW$^+$24) runs on BN256 pairings, while JESS$^+$26 answers a $T$-of-$n$ threshold question with neither linking nor tracing.

\noindent\textbf{Comparison with existing constructions.}
FLLL$^+$21 is the other post-quantum scheme that both links and traces, so it is the one comparison not mediated by a capability gap.
\textsf{PLATRS} signs 3188$\times$ faster (9.886~ms versus 31513.5~ms in its operation-count model), verifies 7231$\times$ faster (3.919~ms versus 28340.9~ms) and returns a 133$\times$ shorter signature (107.3~KiB versus 14257~KiB).
FLLL$^+$21 proves the whole ring in Stern style, whereas \textsf{PLATRS} folds tag and ciphertext into one one-out-of-many relation and delegates the ring to $\kappa$ selector digits. FLLL$^+$21 also opens publicly, while \textsf{PLATRS} keeps the opening with the authority.

\noindent\textbf{What accountability costs.}
Appendix~C deletes one block row of $\mathbf{M}$ at a time: the tag accounts for 11.8\% of the length and 1.54~ms of the signing time, the ciphertext for 41.7\% and 3.33~ms, and the two shares are additive to the byte (53.5\% and 4.84~ms together).

\begin{figure*}[t]
	\centering
	\includegraphics[width=\textwidth]{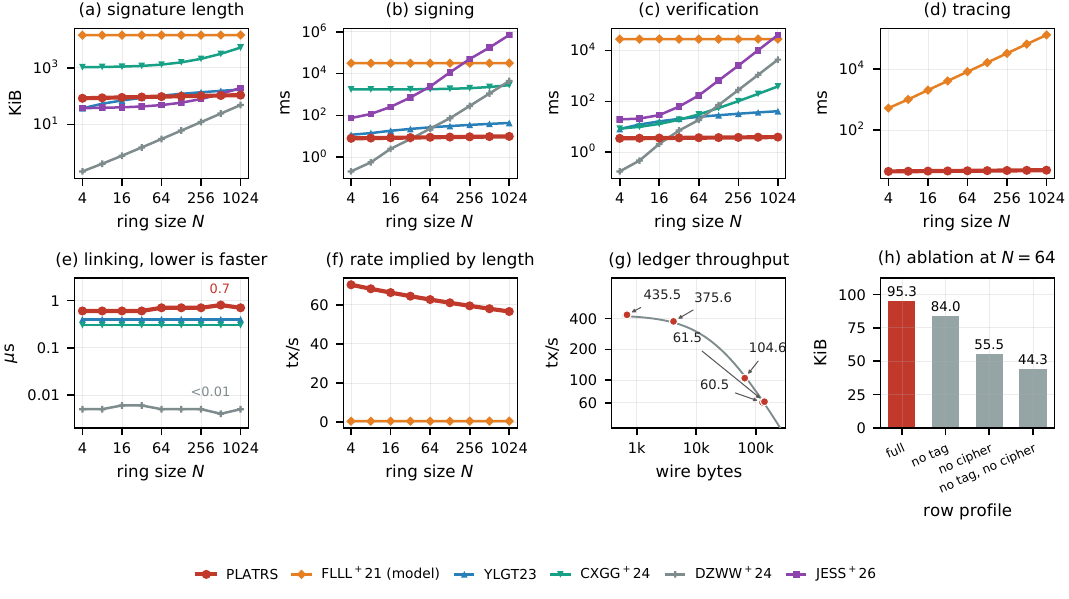}
\caption{\textsf{PLATRS} against the evaluated baselines for $N\in\{4,\dots,1024\}$ (log--log). (a)~Length: \textsf{PLATRS} steps up by one selector digit per doubling of the ring while accumulator schemes stay flat. (b),(c)~Signing and verification: flat in $N$, since the ring enters only through $\kappa$ digits. (d)~Tracing: only the two schemes that open appear. (e)~Linking: a tag comparison, orders of magnitude below one verification. (f)~Commit rate implied by that length through the ledger law of Section~\ref{sec:perf-blockchain}. (g)~Measured commit rate on the Hyperledger Fabric testbed of Section~\ref{sec:perf-blockchain}: one red point per payload from $512$~B to $105$~kB of wire bytes, with the fitted law drawn through them. (h)~Ablation at $N=64$: signature length as each component is removed. Latencies and lengths are measured under one harness; panel~(f) applies the ledger law of Section~\ref{sec:perf-blockchain} to those lengths.}
\label{fig:perf}
\end{figure*}

\noindent\textbf{Against the faster schemes, priced individually.}
YLGT23~\cite{YLGT23} is 4.38$\times$ slower to sign and 10.4$\times$ slower to verify, with a signature 1.57$\times$ longer than ours, but it traces publicly: whoever detects a collision learns the signer, unrestrictedly.
CXGG$^+$24~\cite{FS-LLRS} is 284$\times$ slower to sign and 98.7$\times$ slower to verify, with a signature 47.8$\times$ longer, and has no opening mechanism; its verification is a single linear check over single-byte coefficients.
DZWW$^+$24~\cite{DZWW+24} signs 445$\times$ and verifies 1122$\times$ slower than \textsf{PLATRS} with a signature 0.45$\times$ ours in length, but runs on BN256 pairings and is not post-quantum.
JESS$^+$26~\cite{jagganath2026lotrs} is 1.81$\times$ as long, but \textsf{PLATRS} signs 70866$\times$ and verifies 10456$\times$ faster, and JESS$^+$26 answers a different question.

\subsection{Blockchain Integration}
\label{sec:perf-blockchain}

We deployed the contract on Hyperledger Fabric~v2.2.5 with one endorsing peer and two committing peers, CouchDB and one orderer, on a 16-core Intel i7-11700 with 32~GB of memory, and drove it with Caliper~0.4.0. The endorsement policy names a single organisation, so a transaction is endorsed once and committed twice, and the contract performs no lattice arithmetic: what follows prices the ledger, not the scheme. Blocks are cut at $\texttt{PreferredMaxBytes}=16$~MiB with a $2$~s timeout.

\noindent\textbf{The ledger charges for bytes.}
Over five payload sizes the committed rate follows $T=2.3~\mathrm{ms}+0.105~\mu\mathrm{s}\times\mathrm{wire\ bytes}$, which is panel~(f) of Fig.~\ref{fig:perf}, with residuals under $4\%$ of the per-size medians except the $3{,}104$-byte cell, whose client could not offer the requested load and which therefore bounds that size from above: a transaction that carries nothing still costs 2.3~ms, the proposal, single endorsement, ordering, validation and two state-database commits, and every additional byte costs 0.105~$\mu$s, that is 9.5~MB/s of marginal commit bandwidth. The block size matters only at the large end: a $4$~MiB block holds about thirty signatures and commits the same payload at 44.0\ (5.72~MB/s), against 60.5\ (7.87~MB/s) for the tuned blocks used here.

\noindent\textbf{Client scaling.}
One client reaches 256.6\ registrations per second at $512$~bytes and four reach 403.0\ on the default blocks; eight and sixteen report $305.6$ and $300.9$, but their windows stretch to $63$~s and $71$~s while committed transactions rise from $12.8$k to $21.3$k, so more clients buy work in flight, not a higher rate.

\noindent\textbf{Size-matched comparison.}
Fabric verifies no signature, so two schemes can be compared only by what storing one signature costs. Driving the same contract with payloads equal to each scheme's length at $N=64$ gives 343.4\ submissions per second at $4{,}139$~bytes ($\textsf{LTRS}$-TDSC24), 108.9\ at $65{,}297$~bytes (JESS$^+$26), 60.5\ at $130{,}048$~bytes and 60.1\ at $140{,}288$~bytes (DualRing), all on the line above; these are not runs of those schemes.

\noindent\textbf{A second, client-independent count.}
The peer log agrees. Committing the $65{,}297$-byte payload, the peer records $40$ blocks carrying $3{,}987$ transactions in $41.2$~s, that is $96.9$ per second against Caliper's $108.9$ over its $36.6$~s window; normalising the peer window to Caliper's gives $109.1$, and the two transaction counts agree to $0.2\%$, so the gap is the window definition and not a measurement error. Over the five cells the ledger containers average 17--29\% of the 16-core host, the chaincode container averaging $5.3\%$ of one core, so no cell is arithmetic- or core-bound.

\section{Scope and Deployment}
\label{sec:limitations}

\noindent\textbf{Detection and the client.}
The tag applies to signatures produced by $\mathsf{Sign}$.
A member running modified software can emit a different tag on the same event, and $\mathsf{Link}$ will not match it (Section~\ref{sec:link-alg}).
Tier~2 settles that case by opening the signature and comparing the public opening components, so detection is preventive for conforming clients and retrospective for the rest.

\noindent\textbf{Opening authority.}
$\mathsf{sk}_{\mathsf{TA}}$ decrypts any ciphertext, so authority-only opening fixes which party can open a signature rather than when.
The ledger records every request and its verdict (Section~\ref{sec:ledger}).
A deployment can split $\mathsf{sk}_{\mathsf{TA}}$ across a threshold committee and require a recorded dispute before an opening, which keeps the opening path auditable without changing the primitive; a condition evaluated on hidden values calls for hidden threshold matching, which can be built without pairings~\cite{SilentShare}.

\noindent\textbf{Anonymity and cost.}
CPA is the security target, and Observation~\ref{obs:cca-link} shows that public per-event linking excludes CCA anonymity.
The index-free check adds $2k$ ring elements (7.5~kB, 7.0\% at Set-III) and one linear check; $\pi_{\mathsf{dec}}$ adds one verification to the adjudication.
Length grows by one selector digit per doubling of the ring, and $\mathsf{Judge}$ scans the ring once.
\section{Conclusion}
\label{sec:conclusion}

\textsf{PLATRS} is a lattice-based ring signature with public per-event linkability for honestly generated signatures and authority-only traceability from M-SIS and M-LWE.
Its contribution is the accountability model: a tag any verifier can compare but only the tracing authority can open, so a consortium detects double-endorsement at block-arrival rate without gaining the power to deanonymise members.
Anonymity reduces to M-LWE, the other properties to M-SIS with M-LWE, and tracing failure is bounded analytically.
Tracing is one decryption independent of the ring followed by an $O(N)$ key scan, and linking is one comparison: opening costs 480~$\mu$s and the adjudication that follows adds a re-verification at the recovered index, 4.386~ms on its own at $N=1024$; linking costs 0.7~$\mu$s. The Fabric deployment sustains 60.5\ signature submissions per second at 7.87~MB/s, the commit path rather than the contract setting the rate.

Across the five measured schemes driven on one machine at identical ring sizes, \textsf{PLATRS} ranks 1, 1\ and 2\ of 5\ on signing, verification and length at $N=1024$ (mean rank 1.33); no scheme in the set is cheaper on all three axes at once.
Against the operation-count model of FLLL$^+$21, the closest point in the set on the link-and-trace axis, it is 3188$\times$ faster to sign, 7231$\times$ faster to verify and 133$\times$ more compact while withholding the opening capability FLLL$^+$21 grants to everyone.
Section~\ref{sec:limitations} describes the deployment boundaries of the design.

\bibliographystyle{IEEEtran}
\bibliography{bib/reference}
%\clearpage
\appendices
\section{Algorithms}
\label{app:algorithms}

Algorithms~\ref{alg:sign} and~\ref{alg:verify} make the two layers explicit: one challenge is drawn after both commitments are absorbed, $\pi_{\mathsf{alg}}$ and $\pi_{\mathsf{sel}}$ are transported separately, and verification is two independent checks.

\begin{algorithm}[h]
\footnotesize
\caption{$\mathsf{Sign}(\mathsf{pp},\mathsf{sk}_\pi,M,\mathsf{event},R)$: two layers, one transcript}
\label{alg:sign}
\begin{algorithmic}[1]
\State \textbf{Input:} $M$, $\mathsf{event}$, $R=\{\mathsf{pk}_1,\dots,\mathsf{pk}_N\}$, $\mathsf{sk}_\pi=({\mathbf{s}}_{1,\pi},{\mathbf{s}}_{2,\pi},{\mathbf{e}}_{1,\pi},{\mathbf{e}}_{2,\pi})$
\State \textbf{Output:} $(\sigma,\boldsymbol{\varrho})$ with $\sigma=(\boldsymbol{\tau}_{\mathsf{link}},\mathbf{c},\hat{\mathbf{u}},\pi_{\mathsf{alg}},\pi_{\mathsf{sel}})$, or $\bot$
\State $\mathbf{A}_{\mathsf{event}}\gets H_1(\mathsf{event})$;\quad $\mathbf{e}_{\mathsf{tag}}\gets\mathsf{Expand}(H_2(\mathbf{s}_{2,\pi}\|\mathsf{event}))$
\State $\boldsymbol{\tau}_{\mathsf{link}}\gets\mathbf{A}_{\mathsf{event}}{\mathbf{s}}_{2,\pi}+\mathbf{e}_{\mathsf{tag}}$;\quad $\mathbf{c}_1\gets\mathbf{B}^{\top}\mathbf{r}+\mathbf{e}_3$;\quad $\mathbf{c}_2\gets\mathbf{P}_{\mathsf{TA}}^{\top}\mathbf{r}+\mathbf{e}_4+\mathsf{encode}({\mathbf{P}}_{2,\pi})$
\State $\mathsf{inst}\gets(\mathbf{A},\mathbf{B},\mathbf{A}_{\mathsf{event}},\mathbf{P}_{\mathsf{TA}},R,\boldsymbol{\tau}_{\mathsf{link}},\mathbf{c})$;\quad $\mathbf{w}\gets(\mathbf{s}_{1,\pi},\mathbf{e}_{1,\pi},\mathbf{s}_{2,\pi},\mathbf{e}_{2,\pi},\mathbf{e}_{\mathsf{tag}},\mathbf{r},\mathbf{e}_3,\mathbf{e}_4,\mathbf{t},\mathbf{e}_t)$
\Statex \quad\emph{Algebraic layer}: witness of dimension $O(k+\ell)$, independent of $N$
\State $\mathbf{y}\xleftarrow{\$}$ mask over the blocks of $\mathbf{w}$;\quad $\mathbf{T}\gets\mathbf{M}\mathbf{y}$\Comment{$\mathbf{M}$ of~\eqref{eq:Mwu}}
\Statex \quad\emph{Selection layer}: one-hot digits of $\pi$ in base $b_{\mathsf{sel}}$
\State $b\gets$ base-$b_{\mathsf{sel}}$ one-hot encoding of $\pi$;\quad $a\xleftarrow{\$}$ discrete Gaussian;\quad $B_{\mathsf{bin}}\gets G(r_b,b,\{c_{j,i}\})$
\State $\mathsf{cm}_{\mathsf{sel}}\gets B_{\mathsf{bin,hi}}$;\quad $\boldsymbol{\varrho}\xleftarrow{\$} S_\eta^{2k}$;\quad $\hat{\mathbf{u}}\gets\mathsf{encode}(\mathbf{P}_{1,\pi},\mathbf{P}_{2,\pi})+\mathbf{H}\boldsymbol{\varrho}$ \Comment{commitment to the key images}
\Statex \quad\emph{Challenge and responses}
\State $c\gets H_3(\mathsf{inst},M,\mathbf{c}_1,\mathbf{c}_2,\hat{\mathbf{u}},\mathbf{T},\mathsf{cm}_{\mathsf{sel}})$
\State $\mathbf{z}\gets\mathbf{y}+c\,\mathbf{w}$;\quad $f_{j,i}\gets a_{j,i}+c\,b_{j,i}$;\quad $g_{j,i}\gets f_{j,i}(c-f_{j,i})$;\quad $z_b\gets r_a+c\,r_b$
\If{$\|\mathbf{z}\|_\infty>B$ \textbf{or} a low part exceeds the decomposition slack}
	\State \textbf{restart} from the algebraic layer\Comment{rejection sampling}
\EndIf
\State \Return $(\sigma,\boldsymbol{\varrho})$ with $\sigma=(\boldsymbol{\tau}_{\mathsf{link}},\mathbf{c},\hat{\mathbf{u}},\pi_{\mathsf{alg}},\pi_{\mathsf{sel}})$, $\pi_{\mathsf{alg}}=(\mathbf{T},\mathbf{z})$, $\pi_{\mathsf{sel}}=(\{f_{j,i}\},\mathsf{cm}_{\mathsf{sel}},z_b)$
\end{algorithmic}
\end{algorithm}

\begin{algorithm}[h]
\footnotesize
\caption{$\mathsf{Verify}(\mathsf{pp},M,\mathsf{event},R,\sigma)$: public, index-free}
\label{alg:verify}
\begin{algorithmic}[1]
\State \textbf{Input:} $M$, $\mathsf{event}$, $R$, $\sigma=(\boldsymbol{\tau}_{\mathsf{link}},\mathbf{c},\hat{\mathbf{u}},\pi_{\mathsf{alg}},\pi_{\mathsf{sel}})$
\State \textbf{Output:} $1$ (well formed) or $0$
\State recover $\mathbf{A}_{\mathsf{event}}=H_1(\mathsf{event})$ and $\mathsf{inst}$ as in Algorithm~\ref{alg:sign}
\State parse $\pi_{\mathsf{alg}}=(\mathbf{T},\mathbf{z})$ and $\pi_{\mathsf{sel}}=(\{f_{j,i}\},B_{\mathsf{bin,hi}},z_b)$
\State $c\gets H_3(\mathsf{inst},M,\mathbf{c}_1,\mathbf{c}_2,\hat{\mathbf{u}},\mathbf{T},B_{\mathsf{bin,hi}})$
\Statex \quad\emph{Selection layer: no index, independent of $N$}
\State $f_{j,0}\gets c-\sum_{i\ge1}f_{j,i}$ for every $j$;\quad $\hat A\gets G(z_b,f,g)-2^{K_B}c\,B_{\mathsf{bin,hi}}$
\If{\eqref{eq:selrec} fails \textbf{or} $g_{j,i}\neq f_{j,i}(c-f_{j,i})$ for some $(j,i)$}
	\State \Return $0$\Comment{Lemma~\ref{lem:bin}: this forces $b_{j,i}\in\{0,1\}$}
\EndIf
\If{\eqref{eq:selpoly} has leading coefficient one at more than one index, or at none}
	\State \Return $0$
\EndIf
\Statex \quad\emph{Algebraic layer over the commitment: no index}
\If{$\|\mathbf{z}\|_\infty>B$ \textbf{or} the absorbed rows of~\eqref{eq:relation} fail, i.e.\ $\tilde{\mathbf{M}}\mathbf{z}-c\,\tilde{\mathbf{u}}\neq\tilde{\mathbf{T}}$ over $\tilde{\mathbf{u}}=(\hat{\mathbf{u}},\boldsymbol{\tau}_{\mathsf{link}},\mathbf{c}_1,\mathbf{c}_2)$}
	\State \Return $0$\Comment{rows 1--2 and~\eqref{eq:comcheck} against the committed pair; one extra module product}
\EndIf
\State \Return $1$\Comment{membership is decided at $\mathsf{Confirm}$ and $\mathsf{Judge}$: $\hat{\mathbf{u}}$ must open to a ring entry}
\end{algorithmic}
\end{algorithm}

\section{Proofs}
\label{app:proofs}

Lemma~\ref{lem:tag-prf} is the information-theoretic core: it gives linkability and hiding at once.

\subsection{Auxiliary Lemmas}
\label{app:lemmas}

\begin{lemma}[Pseudorandomness of the tag]
	\label{lem:tag-prf}
	Let $H_2$ and $\mathsf{Expand}$ be modelled as random oracles, so that for each fresh event the vector $\mathbf{e}_{\mathsf{tag}}=\mathsf{Expand}(H_2(\mathbf{s}_2\|\mathsf{event}))$ is uniform over $\{-\eta,\dots,\eta\}^{k\cdot d}$ and independent of everything not derived from $(\mathbf{s}_2,\mathsf{event})$.
	Then for any polynomial number of events the pairs $(\mathbf{A}_{\mathsf{event}},\mathbf{A}_{\mathsf{event}}\mathbf{s}_2+\mathbf{e}_{\mathsf{tag}})$ are jointly indistinguishable from uniform under M-LWE$_{k,\ell,q,\eta}$, and any two of them coincide in $\mathbf{e}_{\mathsf{tag}}$ whenever the same $(\mathbf{s}_2,\mathsf{event})$ recurs.
\end{lemma}

\begin{proof}
	Determinism is immediate, both $\mathbf{e}_{\mathsf{tag}}$ and $\mathbf{A}_{\mathsf{event}}$ being functions of $(\mathbf{s}_2,\mathsf{event})$.
	For a fresh event the oracle output is uniform and independent of $\mathbf{s}_2$, so the tag is exactly an M-LWE sample of $\mathbf{s}_2$; polynomially many such samples of one secret are jointly pseudorandom, since stacking them gives an M-LWE$_{mk,\ell,q}$ instance, and a standard hybrid replaces the tags of the queried events one at a time.
\end{proof}

\begin{lemma}[Key hiding]
\label{lem:key-hiding}
Let $\mathbf{B}\xleftarrow{\$}\mathcal{R}_q^{k\times\ell}$, $\mathbf{s}\xleftarrow{\$} S_\eta^{\ell}$, $\mathbf{e}\xleftarrow{\$}\chi^{k}$ and $\mathbf{P}=\mathbf{B}\mathbf{s}+\mathbf{e}$.
An adversary that sees $\mathbf{P}$ together with polynomially many further M-LWE samples of $\mathbf{s}$ outputs $\hat{\mathbf{s}}\in S_\eta^{\ell}$ with $\|\mathbf{P}-\mathbf{B}\hat{\mathbf{s}}\|_\infty\le\eta$ with probability at most $\mathsf{Adv}^{\mathsf{M\text{-}LWE}}_{k,\ell,q,\eta}+(2\eta+1)^{(\ell+k)d}/q^{kd}$.
\end{lemma}

\begin{proof}
\emph{Step 1 (remove the extra samples).} The polynomially many samples of $\mathbf{s}$ stack into one $\mathsf{M\text{-}LWE}_{k,\ell,q,\eta}$ instance, so replacing them by uniform changes the success probability by at most $\mathsf{Adv}^{\mathsf{M\text{-}LWE}}$, and in the resulting game the view of $\mathcal{A}$ is uniform except for $\mathbf{P}$.

\emph{Step 2 (distinguish).} $\mathcal{B}$ receives $(\mathbf{B}',\mathbf{y}')$, sets $\mathbf{B}=\mathbf{B}'$ and $\mathbf{P}=\mathbf{y}'$, simulates the remaining samples uniformly, and answers ``LWE'' iff $\mathcal{A}$ returns $\hat{\mathbf{s}}\in S_\eta^{\ell}$ with $\|\mathbf{y}'-\mathbf{B}'\hat{\mathbf{s}}\|_\infty\le\eta$.
On an LWE challenge, $\mathbf{y}'=\mathbf{B}'\mathbf{s}+\mathbf{e}$ with $\|\mathbf{e}\|_\infty\le\eta$, so every success of $\mathcal{A}$ is accepted and the acceptance probability is that of the game.
On a uniform $\mathbf{y}'$, acceptance needs $\|\mathbf{y}'-\mathbf{B}'\hat{\mathbf{s}}\|_\infty\le\eta$ for some candidate; for each fixed $\hat{\mathbf{s}}$ this holds with probability $((2\eta+1)/q)^{kd}$ over $\mathbf{y}'$, and a union bound over the $(2\eta+1)^{\ell d}$ elements of $S_\eta^{\ell}$ bounds it by $(2\eta+1)^{(\ell+k)d}/q^{kd}$.
No homogeneous vector is needed here: the tested statement is consistency of a short candidate with a public key, which is exactly what tracing checks.

At $\eta=3$, $q\ge2^{23}$, $k=5$ and $\ell=4$ the bound is $(7^{9}/q^{5})^{d}\le2^{-89d}$.
\end{proof}

\begin{lemma}[Tracing forces exact match]
	\label{lem:trace-exact}
	If a valid $\sigma$ has decrypted key $\tilde{\mathbf{P}}_2$ with $\lVert\tilde{\mathbf{P}}_2-\mathbf{P}_{2,i}\rVert_\infty\le\beta_{\mathsf{trace}}$ for some $i\in[N]$, then except with probability at most $(2\beta_{\mathsf{trace}}+1)^{kd}/q^{kd}+\mathsf{negl}(\lambda)$ the signer's public opening component equals $\mathbf{P}_{2,i}$ exactly.
\end{lemma}

\begin{proof}
	The NIZKPoK gives $\tilde{\mathbf{P}}_2=\mathbf{P}_{2,\pi}+\mathbf{e}_{\mathsf{dec}}$ with $\pi\in[N]$ and $\lVert\mathbf{e}_{\mathsf{dec}}\rVert_\infty\le\beta_{\mathsf{trace}}$.
	If $\pi\neq i$, the two keys are distinct M-LWE samples indistinguishable from independent uniform vectors, so all $kd$ coefficients of their difference lie within $2\beta_{\mathsf{trace}}$ with probability at most $((2\beta_{\mathsf{trace}}+1)/q)^{kd}$ up to the M-LWE advantage.
\end{proof}

\subsection{Correctness (Definition~\ref{def:correct})}
\label{app:correctness}

\begin{proof}
	For an honest signer $\mathbf{e}_{\mathsf{tag}}$ is expanded from $H_2(\mathbf{s}_{2,\pi}\|\mathsf{event})$ with $\lVert\mathbf{e}_{\mathsf{tag}}\rVert_\infty\le\eta$, and $\boldsymbol{\tau}_{\mathsf{link}}=\mathbf{A}_{\mathsf{event}}\mathbf{s}_{2,\pi}+\mathbf{e}_{\mathsf{tag}}$ holds by construction; all other relations hold by honest generation and rejection sampling terminates except negligibly, so verification accepts.
	Per event, $\mathbf{e}_{\mathsf{tag}}$ and $\boldsymbol{\tau}_{\mathsf{link}}$ are deterministic in $(\mathbf{s}_{2,\pi},\mathsf{event})$, so two signatures of one user on one event link.
	Decryption recovers $\mathbf{P}_{2,\pi}$ up to $\beta_{\mathsf{trace}}$ noise except negligibly (Proposition~\ref{prop:trace-bound}) and exactly one member matches (Proposition~\ref{prop:trace-uniqueness}).
\end{proof}

\subsection{Anonymity (Theorem~\ref{thm:anon})}
\label{app:anon}

\begin{proof}
	Let $W_j$ be the event that $\mathcal{A}$ guesses $b$ in game $G_j$.

	\paragraph{Game $G_0$.} The CPA anonymity game: $\Pr[W_0]=\frac12+\mathsf{Adv}^{\mathsf{anon}}_{\mathcal{A}}(\lambda)$.

	\paragraph{Game $G_1$.} The challenge proof $\pi^{\ast}=(\pi_{\mathsf{alg}},\pi_{\mathsf{sel}})$ is replaced by the multi-theorem simulator, which programs $H_3$ after $\mathbf{T}$ and $\mathsf{cm}_{\mathsf{sel}}$ are absorbed: $\mathbf{z}$ is drawn from the response distribution and $\mathbf{T}\gets\mathbf{M}\mathbf{z}-c\,\mathbf{u}$ fixes the algebraic commitment, while $f,g,z_b$ are drawn so that the reconstruction~\eqref{eq:selrec} closes, which needs no opening of the hiding commitment.
	Rejection sampling makes the responses witness independent and $|x|\ll\sigma_a$, so $f=a+xb$ is statistically independent of $b$; hence $|\Pr[W_1]-\Pr[W_0]|\le\mathsf{negl}(\lambda)$.

	\paragraph{Game $G_2$.} $\boldsymbol{\tau}_{\mathsf{link}}^{*}$ becomes uniform $\mathbf{U}\xleftarrow{\$}\mathcal{R}_q^k$.
	By Lemma~\ref{lem:tag-prf}, for a fresh $\mathsf{event}^{*}$ the vector $\mathbf{e}_{\mathsf{tag}}^{*}$ is uniform on $\{-\eta,\dots,\eta\}^{kd}$ and independent of $\mathbf{s}_{2,i_b}$, so the honest tag is an M-LWE sample of $\mathbf{s}_{2,i_b}$; together with $\mathbf{P}_{2,i_b}$ and the tags of the non-challenge signing queries these are polynomially many samples of one secret, jointly pseudorandom by stacking as in Lemma~\ref{lem:tag-prf}. Hence $|\Pr[W_2]-\Pr[W_1]|\le\mathsf{Adv}^{\mathsf{MLWE}}$.

	\paragraph{Game $G_3$.} $\mathbf{c}^{*}$ becomes an encryption of $\mathbf{0}$; IND-CPA of $\Pi_{\mathsf{TA}}$ gives $|\Pr[W_3]-\Pr[W_2]|\le\mathsf{Adv}^{\mathsf{MLWE}}$.

	\paragraph{Conclusion.} In $G_3$ the challenge is independent of $b$, so $\Pr[W_3]=\frac12$; adding the Fiat--Shamir collision term $(q_s+q_h)^2/|\mathcal{C}|$,
	\[
 \mathsf{Adv}^{\mathsf{anon}}_{\mathcal{A}}(\lambda)\le 2\,\mathsf{Adv}^{\mathsf{MLWE}}+\frac{(q_s+q_h)^2}{|\mathcal{C}|}+\mathsf{negl}(\lambda).
 \]
\end{proof}

\subsection{Linkability for Honestly Generated Signatures (Theorem~\ref{thm:link})}
\label{app:link}

\begin{proof}
	\emph{Game $G_0$.} The game of Definition~\ref{def:link-honest}.

	\emph{Game $G_1$.} For $j=1,2$ the signatures are verified by the simulator-extractor of Section~\ref{sec:zkpok}, which returns a witness $w^{(j)}$ and a selector $b^{(j)}$ with isolated index $\pi^{\ast}_j$; by simulation extractability $|\Pr[W_1]-\Pr[W_0]|\le\mathsf{negl}(\lambda)$.

	In $G_1$ extraction follows the two steps of Section~\ref{sec:proof-arch}, and neither step depends on $N$ beyond the selector.
	Step~1 gives $\mathbf{M}\mathbf{w}^{(j)}=\mathbf{u}^{(j)}$ with $\|\mathbf{w}^{(j)}\|_\infty\le B$, hence, in the third block row, $\boldsymbol{\tau}_j=\mathbf{A}_{\mathsf{event}}\mathbf{s}_2^{(j)}+\mathbf{e}_{\mathsf{tag}}^{(j)}$ with $\|\mathbf{e}_{\mathsf{tag}}^{(j)}\|_\infty\le\eta$,
	while the first two rows place $\mathbf{P}_{1,\pi^{\ast}_j}$ and $\mathbf{P}_{2,\pi^{\ast}_j}$ in $\mathbf{u}^{(j)}$.
	Step~2 gives $b^{(j)}$ one-hot at $\pi^{\ast}_j$ over the concatenated keys $Q_i=(\mathbf{P}_{1,i},\mathbf{P}_{2,i})$, so both components in $\mathbf{u}^{(j)}$ belong to the same ring member; this is the only place where the selector and the extracted key are tied together.

	\emph{Condition~(1)} is Definition~\ref{def:correct}(ii): two signatures returned by $\mathsf{Sign}$ for one honest member on one event carry the same $\mathbf{e}_{\mathsf{tag}}$ and the same $\mathbf{A}_{\mathsf{event}}$, hence the same tag.

	\emph{Condition~(2)}. Assume $\mathsf{Link}(\sigma_1,\sigma_2)=\mathsf{linked}$, so $\boldsymbol{\tau}_1=\boldsymbol{\tau}_2$. Subtracting the tag relation,
	\[
 [\mathbf{A}_{\mathsf{event}}\mid\mathbf{I}_k]\cdot\left[\mathbf{s}_2^{(1)}-\mathbf{s}_2^{(2)};\ \mathbf{e}_{\mathsf{tag}}^{(1)}-\mathbf{e}_{\mathsf{tag}}^{(2)}\right]^{\top}=\mathbf{0},
 \]
	so $\mathbf{s}_2^{(1)}\neq\mathbf{s}_2^{(2)}$ yields an M-SIS$_{k,\ell,q,2\eta}$ solution of norm at most $2\eta\le\beta_{\mathsf{sis}}$, the identity block costing only a vanishing advantage; otherwise $\mathbf{e}_{\mathsf{tag}}^{(1)}=\mathbf{e}_{\mathsf{tag}}^{(2)}$ and the two signatures carry the same tag.

	\emph{Case A} (two distinct traced keys, one honest).
	Lemma~\ref{lem:trace-exact} gives $\mathbf{P}_{2,\pi^{(j)}}=\mathbf{B}\mathbf{s}_2^*+\mathbf{e}_2^{(j)}$ exactly, and $\pi^{(1)}=i$ honest gives $\mathbf{B}\mathbf{s}_2^*+\mathbf{e}_2^{(1)}=\mathbf{B}\mathbf{s}_{2,i}+\mathbf{e}_{2,i}$.
	If $\mathbf{s}_2^*\neq\mathbf{s}_{2,i}$ this is an M-SIS$_{k,\ell,q,2\beta}$ solution; otherwise $\mathbf{e}_2^{(1)}=\mathbf{e}_{2,i}$, so either the second signer's secret coincides (probability at most $(2\eta+1)^{-\ell d}$) or the adversary recovered $\mathbf{s}_{2,i}$, excluded by Lemma~\ref{lem:key-hiding}.

	\emph{Case B} (framing an honest user).
	For framed $i$ some signature has $\mathbf{P}_{2,\pi^{(j)}}=\mathbf{P}_{2,i}$ by Lemma~\ref{lem:trace-exact}, hence $[\mathbf{B}\mid\mathbf{I}_k][\mathbf{s}_2^{(j)}-\mathbf{s}_{2,i};\mathbf{e}_2^{(j)}-\mathbf{e}_{2,i}]^{\top}=\mathbf{0}$: a non-zero difference is an M-SIS solution of norm at most $2\beta$, and a zero difference means the honest key was output, excluded by Lemma~\ref{lem:key-hiding} since that user signed at most once.
	Thus $\mathsf{Adv}^{\mathsf{link}}_{\mathcal{A}}\le q_h\big(\mathsf{Adv}^{\mathsf{MSIS}}+\mathsf{Adv}^{\mathsf{MLWE}}+1/|\mathcal{C}|\big)+2\varepsilon_{\mathsf{div}}$.
\end{proof}

\subsection{Traceability (Theorem~\ref{thm:trace})}
\label{app:trace}

\begin{proof}
	\emph{Game $G_0$.} The traceability game of Definition~\ref{def:trace}.

	\emph{Game $G_1$.} The pair is verified by the simulator-extractor, which returns $w^{(j)}$ for $j=1,2$; simulation extractability gives $|\Pr[W_1]-\Pr[W_0]|\le\mathsf{negl}(\lambda)$.
	Step~1 of the extraction places witnesses satisfying $\mathbf{M}\mathbf{w}^{(j)}=\mathbf{u}^{(j)}$: each carries a well-formed committed pair through the absorbed rows, and the ciphertext row encrypts $\mathsf{encode}(\mathbf{P}_2^{(j)})$, the \emph{witness} key; the selector places a declaration $\pi^{\ast}_j$, which plays no role in what follows.

	\emph{Opening behaviour}: decryption recovers the witness key up to $\beta_{\mathsf{trace}}$ except negligibly (Proposition~\ref{prop:trace-bound}), and the match in $R$ is unique (Proposition~\ref{prop:trace-uniqueness}); a witness key outside $R$ matches no member, and $\mathsf{Judge}$ returns $\mathsf{acquit}$.

	\emph{No misattribution}: the recovered key is the witness key $\mathbf{P}_2^{(j)}$ of one of the two signatures, and by the tag equation both share $\mathbf{s}_2^*$ except for the M-SIS event of Theorem~\ref{thm:link}, so they trace to the same key.

	\emph{No framing}: if $\mathsf{pk}^*=\mathsf{pk}_i$ for an honest $i$ that never double-signed, the match gives $\|\mathbf{P}_2^{(j)}-\mathbf{P}_{2,i}\|_\infty\le2\beta_{\mathsf{trace}}$ except negligibly; the key equations at $i$ then give $[\mathbf{B}\mid\mathbf{I}_k\mid-\mathbf{I}_k][\mathbf{s}_2^{(j)}-\mathbf{s}_{2,i};\,\mathbf{e}_2^{(j)}-\mathbf{e}_{2,i};\,\mathbf{P}_2^{(j)}-\mathbf{P}_{2,i}]^{\top}=\mathbf{0}$, so a non-zero difference with $\mathbf{s}_2^{(j)}\neq\mathbf{s}_{2,i}$ is an M-SIS$_{k,\ell,q,\max\{2\beta,2\beta_{\mathsf{trace}}\}}$ solution, and $\mathbf{s}_2^{(j)}=\mathbf{s}_{2,i}$ is excluded by Lemma~\ref{lem:key-hiding}.

	Therefore $\mathsf{Adv}^{\mathsf{trace}}_{\mathcal{A}}\le q_h\big(\mathsf{Adv}^{\mathsf{MSIS}}+\mathsf{Adv}^{\mathsf{MLWE}}+1/|\mathcal{C}|\big)+2\varepsilon_{\mathsf{div}}+\varepsilon_{\mathsf{dec}}$.
\end{proof}

\subsection{Unforgeability (Theorem~\ref{thm:unforgeability})}
\label{app:unforge}

\begin{proof}
	\emph{Game $G_0$.} The unforgeability game of Definition~\ref{def:unforgeability}.

	\emph{Game $G_1$.} The forgery is verified by the simulator-extractor, which returns $w$ and a one-hot selector with index $\pi^{\ast}$; simulation extractability gives $|\Pr[W_1]-\Pr[W_0]|\le\mathsf{negl}(\lambda)$.
	Step~1 of the extraction yields $\mathbf{M}\mathbf{w}=\mathbf{u}$ with $\|\mathbf{w}\|_\infty\le B$, so the absorbed rows give a committed pair $(\mathbf{P}_1,\mathbf{P}_2)$ with $\hat{\mathbf{u}}=\mathsf{encode}(\mathbf{P}_1,\mathbf{P}_2)+\mathbf{H}\boldsymbol{\varrho}'$, $\mathbf{P}_1=\mathbf{A}\mathbf{s}_1+\mathbf{e}_1$ and $\mathbf{P}_2=\mathbf{B}\mathbf{s}_2+\mathbf{e}_2$ for small witnesses; the selector places the declared $\pi^{\ast}$, which is not needed for the bridge.
	$\mathsf{Confirm}(i,\sigma,\boldsymbol{\varrho})=1$ for the named $i$ gives $\hat{\mathbf{u}}=\mathsf{encode}(\mathbf{P}_{1,i},\mathbf{P}_{2,i})+\mathbf{H}\boldsymbol{\varrho}$.  Subtracting the two openings leaves $\mathsf{encode}(\mathbf{P}_{1,i},\mathbf{P}_{2,i})-\mathsf{encode}(\mathbf{P}_1,\mathbf{P}_2)=\mathbf{H}(\boldsymbol{\varrho}'-\boldsymbol{\varrho})$: either the pairs coincide, or the difference is a nonzero M-SIS$_{k,\cdot,q,2\eta}$ solution for $\mathbf{H}$.  In the coinciding case, $i$ is honest and subtracting the honest relation $\mathbf{P}_{1,i}=\mathbf{A}\mathbf{s}_{1,i}+\mathbf{e}_{1,i}$ gives $[\mathbf{A}\mid\mathbf{I}_k][\mathbf{s}_1-\mathbf{s}_{1,i};\mathbf{e}_1-\mathbf{e}_{1,i}]^{\top}=\mathbf{0}$: a non-zero difference is an M-SIS$_{k,\ell,q,2\beta}$ solution, a zero difference means the honest secret was output, excluded by Lemma~\ref{lem:key-hiding}.
	Thus $\mathsf{Adv}^{\mathsf{unforge}}_{\mathcal{A}}\le q_h\big(\mathsf{Adv}^{\mathsf{MSIS}_{2\beta}}+\mathsf{Adv}^{\mathsf{MSIS}_{2\eta}}+1/|\mathcal{C}|\big)+2\varepsilon_{\mathsf{div}}$.
\end{proof}

\subsection{Simulation extractability}
\label{app:sim-ext}

\noindent\textbf{The relation.}
A statement is a pair $(\mathsf{inst},\mathbf{u})$ with $\mathsf{inst}=(\mathbf{A},\mathbf{B},\mathbf{A}_{\mathsf{event}},\mathbf{P}_{\mathsf{TA}},\mathcal{C})$ and $\mathbf{u}=(\hat{\mathbf{u}},\boldsymbol{\tau},\mathbf{c}_1,\mathbf{c}_2,\mathbf{c}_{\mathsf{sel}})$, and a witness is $\mathbf{w}=(\mathbf{P}_1,\mathbf{P}_2,\boldsymbol{\varrho},\mathbf{s}_1,\mathbf{s}_2,\mathbf{e}_1,\mathbf{e}_2,\mathbf{e}_{\mathsf{tag}},\mathbf{e}_3,\mathbf{e}_4,\mathbf{r},b,\mathbf{r}_b,\mathbf{y},\mathbf{a})$ with
\[
\mathcal{R}=\bigl\{(\mathsf{inst},\mathbf{u};\mathbf{w}):\ \mathbf{M}\mathbf{w}=\mathbf{u},\ \|\mathbf{w}\|_\infty\le B\bigr\}.
\]
Table~\ref{tab:relation} lists the block rows of $\mathbf{M}$: each is one relation of the construction, the response column is the corresponding block of $\mathbf{z}=\mathbf{y}+c\mathbf{w}$, and only the last row is answered in the selector layer.

\begin{table}[t]
\centering\scriptsize
\setlength{\tabcolsep}{3pt}
\caption{Blocks of the proved relation and the response that opens each block.}
\label{tab:relation}
\begin{tabular}{@{}p{0.40\columnwidth}p{0.17\columnwidth}p{0.39\columnwidth}@{}}
\toprule
Relation & Witness & Response\\
\midrule
$\mathbf{P}_1=\mathbf{A}\mathbf{s}_1+\mathbf{e}_1$ & $\mathbf{s}_1,\mathbf{e}_1$ & $\mathbf{z}_1=\mathbf{y}_1+c\mathbf{s}_1$\\
$\mathbf{P}_2=\mathbf{B}\mathbf{s}_2+\mathbf{e}_2$ & $\mathbf{s}_2,\mathbf{e}_2$ & $\mathbf{z}_2=\mathbf{y}_2+c\mathbf{s}_2$\\
$\boldsymbol{\tau}=\mathbf{A}_{\mathsf{event}}\mathbf{s}_2+\mathbf{e}_{\mathsf{tag}}$ & $\mathbf{s}_2,\mathbf{e}_{\mathsf{tag}}$ & $\mathbf{z}_3=\mathbf{y}_2+c\mathbf{s}_2$,\newline $\mathbf{z}_{t}=\mathbf{y}_{t}+c\mathbf{e}_{\mathsf{tag}}$\\
$\mathbf{c}_1=\mathbf{B}^{\top}\mathbf{r}+\mathbf{e}_3$ & $\mathbf{r},\mathbf{e}_3$ & $\mathbf{z}_{r}=\mathbf{y}_{r}+c\mathbf{r}$,\newline $\mathbf{z}_{3}=\mathbf{y}_{3}+c\mathbf{e}_3$\\
$\mathbf{c}_2=\mathbf{P}_{\mathsf{TA}}^{\top}\mathbf{r}+\mathbf{e}_4$ & $\mathbf{r},\mathbf{e}_4$ & $\mathbf{z}_{r}$,\newline $\mathbf{z}_{4}=\mathbf{y}_{4}+c\mathbf{e}_4$\\
$\quad{}+\mathsf{encode}(\mathbf{P}_2)$ & & \\
$\mathbf{c}_{\mathsf{sel}}=\mathsf{Com}_{\mathsf{bin}}(b;\{c_{b,i}\})$ & $b,\mathbf{r}_b$ & $f_j=a_j+xb_j$,\newline $g_j=f_j(x-f_j)$\\
\bottomrule
\end{tabular}
\end{table}

\noindent\textbf{Absorption.}
Rows 1--2 absorb the committed pair through $\mathbf{P}_1=\hat{\mathbf{u}}_1-\mathbf{H}_1\boldsymbol{\varrho}$ and $\mathbf{P}_2=\hat{\mathbf{u}}_2-\mathbf{H}_2\boldsymbol{\varrho}$, with $\hat{\mathbf{u}}=\hat{\mathbf{u}}_1\oplus\hat{\mathbf{u}}_2$ split along the coordinate blocks of $\mathsf{encode}$, and the $\mathsf{encode}(\mathbf{P}_2)$ term of the ciphertext row is absorbed the same way; the statement $\mathbf{u}=(\hat{\mathbf{u}},\boldsymbol{\tau},\mathbf{c}_1,\mathbf{c}_2,\mathbf{c}_{\mathsf{sel}})$ therefore contains no key in the clear and every block row is evaluable from $(\mathbf{z},\mathbf{u})$ alone.  At a named $i$ with $\boldsymbol{\varrho}$, the opening $\hat{\mathbf{u}}=\mathsf{encode}(\mathbf{P}_{1,i},\mathbf{P}_{2,i})+\mathbf{H}\boldsymbol{\varrho}$ collapses the absorbed rows onto the rows at $\mathbf{u}_i$, which is what $\mathsf{Confirm}$ runs.

\noindent\textbf{Simulation.}
The simulator samples $c\in\mathcal{C}$ uniformly, draws the response from its marginal distribution and sets $\mathbf{T}\leftarrow\mathbf{M}\mathbf{z}-\mathbf{u}$ (Appendix~\ref{app:anon}), so its transcripts satisfy the verification equations exactly and carry no information on $\mathbf{w}$.
This is multi-theorem: each proof programs $H_3$ at its own point, so the number of programmed points is $q_s+1$, and none of them is a forgery, which the game excludes.

\noindent\textbf{Extraction.}
Let $\mathcal{A}$ output $(\mathsf{inst},\pi)$, accepted by $\mathsf{Verify}$ and not returned by $\mathsf{Sign}$, with probability $\varepsilon$, and let $c^{\ast}=H_3(\mathsf{inst},M,\mathbf{com}^{\ast})$ be its challenge.
Since $\mathbf{com}^{\ast}$ is fixed before $c^{\ast}$ is queried, $c^{\ast}$ is uniform in $\mathcal{C}$ and independent of $\mathbf{com}^{\ast}$.
Rerunning $\mathcal{A}$ on the same coins while answering the $c^{\ast}$ query with a fresh $c'\in\mathcal{C}$ yields, by the forking lemma, a second accepting transcript on $(\mathsf{inst},M,\mathbf{com}^{\ast})$ with $c'\neq c^{\ast}$ with probability at least $\varepsilon/q_h-1/|\mathcal{C}|$.
The two transcripts satisfy $\mathbf{M}\mathbf{z}=\mathbf{T}+c^{\ast}\mathbf{u}$ and $\mathbf{M}\mathbf{z}'=\mathbf{T}+c'\mathbf{u}$ with the same first move $\mathbf{T}$, so $\mathbf{M}(\mathbf{z}-\mathbf{z}')=(c^{\ast}-c')\mathbf{u}$ and, dividing by the unit $c^{\ast}-c'$ (Definition~\ref{def:unitsep}), $\mathbf{w}=(\mathbf{z}-\mathbf{z}')/(c^{\ast}-c')$ satisfies $\mathbf{M}\mathbf{w}=\mathbf{u}$ with $\|\mathbf{w}\|_\infty\le2B$.
The selector is read from the same two transcripts; when its first move has three invertible differences, Lemma~\ref{lem:sel-extract} applies and the loss is the $2\,\mathsf{lf}(\varepsilon)+2\varepsilon_{\mathsf{div}}$ already carried in the bounds.

\begin{proposition}[Simulation extractability]
\label{prop:sim-ext}
Let $\mathcal{A}$ make $q_h$ queries to $H_3$ and $q_s$ queries to $\mathsf{Sign}$, and let it output $(\mathsf{inst},\pi)$ accepted by $\mathsf{Verify}$ although $\mathsf{Sign}$ never returned it, with probability $\varepsilon$.
Then the extractor above returns $\mathbf{w}$ with $\mathbf{M}\mathbf{w}=\mathbf{u}$, $\|\mathbf{w}\|_\infty\le2B$, and a one-hot index $\pi^{\ast}$, with probability at least $\varepsilon/q_h-1/|\mathcal{C}|-\varepsilon_{\mathsf{div}}$.
\end{proposition}
\begin{proof}
The failure events are a failed fork, bounded by $1-\varepsilon/q_h+1/|\mathcal{C}|$, a non-unit challenge difference, bounded by $\varepsilon_{\mathsf{div}}$, and the selector's second fork, carried as $\mathsf{lf}(\varepsilon)$ in the theorem bounds.
Simulated proofs do not enter: the forgery is required not to be one, and programming $H_3$ only changes answers at points chosen after the commitment, which leaves $c^{\ast}$ uniform and the two transcripts independent of $\mathbf{w}$.
\end{proof}

\subsection{Tracing Noise (Proposition~\ref{prop:trace-bound})}
\label{app:noise}

\begin{proposition}[Tracing noise bound]
	\label{prop:trace-bound}
	For Set-III with $d=256$, $k=5$, $\ell=6$, $\eta=3$ and parameters as in Table~\ref{tab:params}, with $\chi$ the signed-sum distribution of variance $v_\chi=\eta$ and $S_\eta$ uniform of variance $v_S=\eta(\eta+1)/3$,
	\[
 \Pr\!\left[\|\mathbf{E}_{\mathsf{TA}}^{\top}\mathbf{r}+\mathbf{e}_4-\mathbf{S}_{\mathsf{TA}}^{\top}\mathbf{e}_3\|_\infty>\beta_{\mathsf{trace}}\right]\le kd\cdot\exp\!\left(-\frac{\beta_{\mathsf{trace}}^2}{2\sigma_{\mathsf{dec}}^2}\right),
 \]
	with $\sigma_{\mathsf{dec}}^2=v_\chi+d\,v_\chi v_S(k+\ell)=33{,}795$; for $\beta_{\mathsf{trace}}=1{,}296$ the bound is below $2.1\times10^{-8}$.
\end{proposition}

\begin{proof}
	Each coefficient of $\mathbf{E}_{\mathsf{TA}}^{\top}\mathbf{r}$ is a sum of $k$ negacyclic convolutions and that of $\mathbf{S}_{\mathsf{TA}}^{\top}\mathbf{e}_3$ a sum of $\ell$, while $\mathbf{e}_4$ contributes one independent term.
	The $d$ products entering one output coefficient of a convolution are independent, pairing an error coefficient of variance $v_\chi$ with one of variance $v_S$, so each convolution coefficient has variance $d\,v_\chi v_S$; the $k+\ell$ convolutions and the direct term are independent, hence $\sigma_{\mathsf{dec}}^2=v_\chi+d\,v_\chi v_S(k+\ell)=33{,}795$.
	The $\chi$ sampler is a sum of $\eta$ independent $\pm1$ terms, so it is sub-Gaussian, and a union bound over the $kd$ coefficients gives the stated tail.
	The worst-case bound is deliberately not used: $\beta_{\mathsf{trace}}$ is fixed from the simulated maximum (Section~\ref{sec:noise}) and this tail certifies the gap.
\end{proof}

\subsection{Tracing Uniqueness (Proposition~\ref{prop:trace-uniqueness})}
\label{app:uniqueness}

\begin{proposition}[Tracing uniqueness]
	\label{prop:trace-uniqueness}
	For a valid signature of $\pi$ and every $j\neq\pi$ in the ring, the probability that $\tilde{\mathbf{P}}_2$ lies within $\beta_{\mathsf{trace}}$ of $\mathbf{P}_{2,j}$ is at most $\left(\frac{2\beta_{\mathsf{trace}}+1}{q}\right)^{kd}+\mathsf{Adv}^{\mathsf{MLWE}}$.
\end{proposition}

\begin{proof}
	For $j\neq\pi$, $\mathbf{P}_{2,j}$ is indistinguishable from uniform, and in the uniform case all $kd$ independent coefficients falling within $\pm\beta_{\mathsf{trace}}$ of $\tilde{\mathbf{P}}_2$ has probability at most $\left(\frac{2\beta_{\mathsf{trace}}+1}{q}\right)^{kd}$; a union bound over $N-1$ non-signers gives $(N-1)(\frac{2\beta_{\mathsf{trace}}+1}{q})^{kd}$, negligible for Table~\ref{tab:params} and practical rings, plus the additive $\mathsf{Adv}^{\mathsf{MLWE}}$ term.
\end{proof}

\subsection{Parameters and Noise}
\label{app:params}
\label{sec:parameters}

\noindent\textbf{Design rationale.}

Four requirements bind the parameters: both M-LWE and M-SIS reach $\lambda=128$ bits; tracing noise stays below $\beta_{\mathsf{trace}}$ (Appendix~B-G); $\beta_{\mathsf{zk}}$ keeps rejection sampling rare while preserving M-SIS hardness for witnesses that include $\mathbf{e}_{\mathsf{tag}}$; and $d,q$ take the smallest values meeting these constraints.
Sets~I and~II are prototypes and Sets~III--IV target deployment; the paper uses Set-III ($d=256$, $q\approx2^{23}$, $(k,\ell)=(5,6)$, $\eta=3$, $\beta_{\mathsf{zk}}=500$, $\beta_{\mathsf{trace}}=1{,}296$).

\noindent\textbf{Security estimation.}
\label{sec:estimator}
Concrete security is estimated with the Lattice Estimator~\cite{Albrecht2018} (commit \texttt{3e48ef4}).
The module instance is modelled as plain LWE with $n=d\ell$, $m=dk$ under the \texttt{usvp}, \texttt{bdd}, \texttt{dual} and \texttt{dual\_hybrid} attacks: Set-III reaches $222$ bits and Set-IV $513$; M-SIS uses $n=d$, $m=d(k+\ell)$ and bound $2\beta_{\mathsf{zk}}$ and gives $202$ and $395$, while Sets~I and~II fall below $128$ bits and must not be deployed.
The tag noise is a sample of $\chi$: $\mathsf{Expand}$ outputs the signed sum of $\eta$ signs derived from $H_2$, so the tag is an M-LWE sample with the noise law of $\mathbf{e}_1,\dots,\mathbf{e}_4$ and inherits the estimate.

\noindent\textbf{Tracing noise.}
\label{sec:noise}
Tracing recovers $\mathbf{P}_{2,\pi}$ up to $\mathbf{e}_{\mathsf{dec}}=\mathbf{E}_{\mathsf{TA}}^{\top}\mathbf{r}+\mathbf{e}_4-\mathbf{S}_{\mathsf{TA}}^{\top}\mathbf{e}_3$, whose variance is $\sigma_{\mathsf{dec}}^2=v_\chi+d\,v_\chi v_S(k+\ell)$ with $v_\chi=\eta$ and $v_S=\eta(\eta+1)/3$ (Appendix~B-G); for Set-III $\sigma_{\mathsf{dec}}\approx184$, and $3.8\times10^5$ simulated coefficients gave $\sigma=183.5$, within $0.2\%$, with a maximum of $864$.
We set $\beta_{\mathsf{trace}}=1{,}296$, $1.5\times$ that maximum and well inside the $2.1\times10^{-8}$ tail of Proposition~\ref{prop:trace-bound}; Proposition~\ref{prop:trace-uniqueness} bounds uniqueness failure independently of $N$, and the union bound over the ring multiplies it by $N$, negligible at every measured size.

\begin{table}[t]
\centering
\caption{Parameter sets.}
\label{tab:params}

\setlength{\tabcolsep}{3pt}
\begin{tabular}{@{}lccccccc@{}}
\toprule
Set & $d$ & $q$ & $k$ & $\ell$ & $\eta$ & M-LWE & M-SIS \\
\midrule
Set-I   & 64  & $2^{16}+1$        & 2 & 3 & 1 & 41  & 62  \\
Set-II  & 128 & $2^{23}-2^{13}+1$ & 4 & 5 & 2 & 89  & 118 \\
Set-III & 256 & $2^{23}-2^{13}+1$ & 5 & 6 & 3 & 222 & 202 \\
Set-IV  & 512 & $2^{23}-2^{13}+1$ & 6 & 7 & 4 & 513 & 395 \\
\bottomrule
\end{tabular}
\par\vspace{3pt}
\parbox{\columnwidth}{\footnotesize M-LWE and M-SIS columns are classical security estimates from the Lattice Estimator. Sets~I and~II are prototyping configurations below 128 bits and are not for deployment.}
\end{table}

\subsection{CPA Anonymity and the Limits of Public Linking}\label{app:cpa}
\label{sec:cpa-vs-cca}

We prove CPA anonymity (Definition~\ref{def:anon}) rather than CCA, for a structural reason: public per-event linkability is a \emph{public predicate}, since the tag of any signature is comparable with that of any other on the same event, and an adversary holding one signature from a challenge user on the challenge event can therefore test equality with the challenge signature itself; no simulator can hide which member signed, because the distinguishing test needs no secret.

\begin{observation}
\label{obs:cca-link}
Let a ring signature scheme be publicly linkable for per-event tags, in the sense that a public deterministic algorithm decides whether two signatures on the same event share a signer.
Then no such scheme satisfies CCA anonymity with respect to a signing oracle that answers queries on the challenge event for challenge users.
\end{observation}

\noindent
\emph{Argument.}
The CCA experiment gives $\mathcal{A}$ a challenge signature $\sigma^\ast$ on event $\mathsf{ev}$ from one of two challenge users and permits it to obtain a signature $\sigma'$ by either challenge user on $\mathsf{ev}$; linkability being public, $\mathcal{A}$ computes the linking decision on $(\sigma^\ast,\sigma')$, and the outcome reveals whether the challenge signer produced $\sigma'$, exactly the bit to be guessed. In game terms, the two games of the CCA experiment differ on an event of probability one for a single oracle query, so no hop can hide it.

What the alternative would need is a tag comparable by the public yet unlinkable under a signing oracle on the same event, which Observation~\ref{obs:cca-link} rules out.

\section{Component Ablation}
\label{app:ablation}

\begin{table}[h!]
\centering
\caption{Ablation at $N=64$: each row deletes one component from the same code path, and all five rows come from one ablation run.}
\label{tab:ablation}
\footnotesize
\setlength{\tabcolsep}{3pt}
\begin{tabular*}{\columnwidth}{@{\extracolsep{\fill}}lccccl@{}}
\toprule
Variant & size & sign & verify & $\Delta$size & lost \\
        & (KiB) & (ms) & (ms) & (KiB) & \\
\midrule
\textsf{PLATRS}          & 95.3 & 8.87 & 3.70 & ---     & --- \\
w/o tag                  & 84.0 & 7.32 & 3.05 & $-11.3$ & linking \\
w/o ciphertext           & 55.5 & 5.54 & 2.28 & $-39.8$ & tracing \\
w/o tag \& ciphertext    & 44.3 & 4.03 & 1.64 & $-51.0$ & both \\
w/o digit selector       & 3785 & 106.5 & 92.5 & $+3690$ & $\log N$ size \\
\bottomrule
\end{tabular*}
\par\vspace{2pt}
{\raggedright\footnotesize Sizes in KiB; timings are medians of sixty runs of that driver, single-threaded, on the machine of Section~VII.\par}
\end{table}

Table~\ref{tab:ablation} recomputes \textsf{PLATRS} with one block row of $\mathbf{M}$ removed, so every difference is the price of that component rather than a comparison between implementations: deleting the tag row removes what $\mathsf{Link}$ compares, deleting the two ciphertext rows removes what $\mathsf{Open}$ decrypts, and the last row replaces the $\kappa$-digit selector with the flat $N$-branch OR of Section~V-E.

No component is redundant: sizes close to the byte and times add up within noise, and the selector is what makes accountability affordable: a flat OR would put tag and ciphertext on a $39.7\times$ longer transcript.

\end{document}